\documentclass[12pt,preprint]{aastex}

\slugcomment{Submitted to Astrophyisical Journal}

\shorttitle{Subaru IR Spectroscopy of HH Driving Sources}
\shortauthors{Takami et al.}

\begin{document}


\title{Subaru IR Echelle Spectroscopy of Herbig-Haro Driving Sources I. H$_2$ and [Fe II] Emission\footnote{Based on data collected at the Subaru Telescope, which is operated by National Astronomical Observatory of Japan}}


\author{M. Takami\altaffilmark{1,2},
        A. Chrysostomou\altaffilmark{2},
        T.P. Ray \altaffilmark{3},
        C.J. Davis \altaffilmark{4},
        W.R.F. Dent  \altaffilmark{5},
        J. Bailey\altaffilmark{6},
        M. Tamura\altaffilmark{7},
        H. Terada\altaffilmark{1},
        T.S. Pyo\altaffilmark{1}}
\email{mtakami@subaru.naoj.org}


\altaffiltext{1}{Subaru Telescope, 650 North A'ohoku Place, Hilo, Hawaii
96720, USA}
\altaffiltext{2}{Centre for Astrophysics Research, University of
           Hertfordshire, Hatfield, HERTS AL10 9AB, UK}
\altaffiltext{3}{School of Cosmic Physics, Dublin Institute for Advanced Studies, 5 Merrion Square, Dublin 2, Ireland}
\altaffiltext{4}{Joint Astronomy Centre, 660 North A'ohoku Place, University Park, Hilo, Hawaii 96720, USA}
\altaffiltext{5}{UK Astronomy Technology Centre, Royal Observatory, Blackford Hill, Edinburgh EH9 3HJ, UK}
\altaffiltext{6}{Australian Centre for Astrobiology, Macquarie University, NSW 2109, Australia}
\altaffiltext{7}{National Astronomical Observatory of Japan, Osawa, Mitaka, Tokyo 181-8588, Japan}


\begin{abstract}
We present infrared echelle spectroscopy of three Herbig-Haro (HH)
driving sources (SVS 13, B5-IRS~1 and HH~34~IRS) using Subaru-IRCS.
The large diameter of the telescope and wide spectral coverage of
the spectrograph allowed us to detect several H$_2$ and [Fe II]
lines in the $H$- and $K$-bands. These include H$_2$ lines arising
from $v$=1--3 and $J$=1--11, and [Fe II] lines with upper level
energies of $E/k=1.1-2.7\times10^4$~K. For all objects the outflow
is found to have two velocity components: (1) a high-velocity ($-$70
to $-$130~km\,s$^{-1}$) component (HVC), seen in [Fe II] or H$_2$
emission and associated with a collimated jet; and (2) a
low-velocity ($-$10 to $-$30~km\,s$^{-1}$) component (LVC), which is
seen in H$_2$ emission only and is spatially more compact. Such a
kinematic structure resembles optical forbidden emission line
outflows associated with classical T Tauri stars, whereas the
presence of H$_2$ emission reflects the low-excitation nature of the
outflowing gas close to these protostars. The observed H$_2$ flux
ratios indicate a temperature of $2-3\times10^3$~K, and a gas
density of $10^5$~cm$^{-3}$ or more, supporting shocks as the
heating mechanism. B5-IRS~1 exhibits faint extended emission
associated with the H$_2$-LVC, in which the radial velocity slowly
increases with distance from the protostar (by $\sim
20$~km\,s$^{-1}$ at $\sim$~500~AU). This is explained as warm
molecular gas entrained by an unseen wide-angled wind. The [Fe II]
flux ratios indicate electron densities to be $\sim$10$^4$~cm$^{-3}$
or greater, similar to forbidden line outflows associated with
classical T Tauri stars. Finally the kinematic structure of the [Fe
II] emission associated with the base of the B5-IRS~1 and HH~34~IRS
outflows is shown to support disk-wind models.
\end{abstract}





\section{Introduction}
Mass accretion and ejection are fundamental processes for the
evolution of young stellar objects (YSOs). Understanding their
mechanisms, and also physical conditions which determine stellar
masses, is one of the major challenges of contemporary astrophysics.
Mass accretion cannot proceed without removing angular momentum from
the surrounding material. Theories predict that outflowing gas plays
an important role in this process (see e.g. Blandford \& Payne 1982;
Shu et al. 2000; K\"onigl \& Pudritz 2000). However, there are two
crucial problems which hinder the investigation of this scenario
through observations. One is the limited spatial resolution of
telescopes, which has not been sufficient to fully resolve the
jet/wind launching region, i.e. where the angular momentum transfer
may occur. The other is the fact that such a region is often deeply
embedded within a massive envelope, making it difficult to observe
at optical to near-IR wavelengths.

Even so, classical T-Tauri stars, a class of low-mass YSOs, are
directly observable at optical-UV wavelengths, allowing the study of
mass accretion and ejection to within 100~AU of the central star for
nearby sources. Studies of UV excess continuum and permitted
emission lines strongly suggest that stellar bipolar magnetic fields
are coupled to the inner edge of the disk at a few stellar radii,
regulating stellar rotation to well below the break-up velocity
(see, e.g., Calvet et al. 2000 and Najita et al. 2000 for reviews).
Disk material seems to lose a significant amount of angular momentum
at the inner edge of the disk, before falling down magnetic field
lines and being shocked near the stellar surface. Such
``magnetospheric accretion flow'' is indeed observed as redshifted
absorption in permitted lines (see, e.g., Najita et al. 2000). At
the same time, a jet and/or wind seems to be launched within a few
AU of the disk or at the inner disk edge. A physical link between
these two phenomena is indicated by the fact that disk accretion
rates, measured from the veiling continuum, are correlated with mass
outflow signatures such as forbidden line luminosities (Hartigan,
Edwards, \& Ghandour 1995; Calvet 1997). Using the {\it Hubble Space
Telescope (HST)} Bacciotti et al. (2002), Coffey et al. (2004) and
Woitas et al. (2005) show that the internal motion of forbidden line
outflows is consistent with flow rotation. These results agree with
the scenario that the flow removes angular momentum from the disk,
thereby allowing mass accretion to occur. Studying the kinematics of
outflowing gas at AU scales, which cannot be directly resolved by
the {\it HST} or adaptive optics with 8-10 m telescopes, could
provide more crucial information for understanding their ejection.
This spatial scale is being explored using the technique of
spectro-astrometry which allows the spatial structure of
emission/absorption features to be studied down to milliarcsecond
scales (e.g., Bailey 1998; Takami et al. 2001; Takami, Bailey \&
Chrysostomou 2003; Whelan et al. 2004).

However, there is growing evidence that the stellar mass has almost
been determined at younger evolutionary phases (Class 0--I phases),
and mass accretion at the T Tauri phase is responsible for only a
tiny fraction of the entire stellar mass (of order 1\% --- see
Calvet et al. 2000). It is often assumed, but not confirmed, that
disk accretion within 100~AU of Class I protostars occurs by the
same mechanism as for T Tauri stars. Br$\gamma$ line profiles
observed towards Class I protostars suggest that these stars are
associated with magnetospheric accretion columns like T Tauri stars
(see e.g., Davis et al. 2001). However, measured Br$\gamma$
luminosities suggest that steady magnetospheric mass accretion can
explain only a few percent of the total stellar mass, requiring
episodic and violent mass accretion (see Calvet et al. 2000).

Class 0--I protostars also exhibit signatures of energetic mass
ejection as optical jets and molecular bipolar outflows (see e.g.,
Reipurth \& Bally 2001 and Richer et al. 2000 for reviews). While
the extended parts of these outflows have been observed over
decades, kinematics of outflowing gas within a few hundred AU has
just begun to be revealed by near-IR high-resolution spectroscopy
and spectro-imaging. Davis et al. (2001) observed H$_2$ 2.122~$\mu$m
emission at several HH driving sources, showing the presence of
Molecular Hydrogen Emission Line (MHEL) regions, analogous to the
forbidden emission line (FEL) regions associated with T Tauri stars.
The H$_2$ emission shows a variety of profiles with multiple
velocity peaks similar to FELs in T-Tauri stars. The kinematics of
the H$_2$ 2.122~$\mu$m emission differs from that of the [Fe II]
1.644~$\mu$m emission, revealing complicated kinematic structure
(Davis et al. 2003). At the same time Fabry-Perot observations
(Davis et al. 2002) show the presence of small scale jets traced by
H$_2$ emission, analogous to forbidden line jets associated with
classical T Tauri stars. Evidence for H$_2$ emission from cavity
walls is also seen for a few objects, suggesting the presence of a
wide-angled wind.

Detailed studies of mass ejection in these objects could lead us to
understand the mechanism of mass accretion, which is responsible for
determining the stellar mass. We thus performed an extensive study
of MHEL regions and [Fe II] emission line regions associated with
three HH driving sources using Subaru-IRCS.

The rest of the paper is organised as follows. In \S 2 we describe
the observations and data reduction methods. In \S 3 we present the
observed kinematics seen in the H$_2$ 2.122~$\mu$m and [Fe II]
1.644~$\mu$m emission, the brightest H$_2$ and [Fe II] emission in
our spectra. In \S 4 we derive the extinction, density and
temperature inferred from H$_2$ and [Fe II] line ratios. In \S 5 we
compare our results with forbidden line outflows associated with T
Tauri stars, and investigate the mechanism of driving jets/winds for
low-mass YSOs. In Paper II we show the results for other emission
lines including CO and atomic lines.


\section{Observations}
Observations were made on 2002 November 25 at the SUBARU 8.2-m
telescope using the Infrared Camera and Spectrograph IRCS (Tokunaga
et al. 1998; Kobayashi et al. 2000). The echelle grating mode with a
0.3\arcsec\, wide slit provides a spectral resolution of
1.1$\times$10$^4$, corresponding to a velocity resolution of
28~km\,s$^{-1}$. The pixel scale of 0.06\arcsec\, provides better
than Nyquist sampling of the seeing ($\sim$~0.6\arcsec), thereby
allowing us to investigate spatial structures on sub-arcsecond
scales.

Table \ref{tbl-targets}  shows the log of the observations. Three
targets, SVS 13, B5-IRS~1 and HH~34~IRS, were selected from Davis et
al. (2001, 2003) based on the strength of the H$_2$ and [Fe II]
emission in their proximity. Details of the individual targets are
described in Appendix A. Spectra in $H$- and $K$-bands were obtained
along the jet axis for all objects, and $K$-band spectra
perpendicular to the jet axis were also obtained for SVS 13. These
cross-dispersed spectra in $H$- and $K$-bands cover 1.52--1.78
$\mu$m (orders 37--32) and 1.96--2.45~$\mu$m (orders 28--23),
respectively.
Thus they include a number of [Fe II] and H$_2$ lines, thereby
allowing for accurate measurements of their relative fluxes. At each
slit position, object-sky-object sequences were repeated each with
an exposure time of 120--180 seconds. Spectra at opposite slit
angles ($\Delta$P.A. = 180$^\circ$) were also obtained to remove bad
pixels. In addition to the targets, A-type bright standards were
observed at similar airmasses to correct for telluric absorption.
The flat fields were made by combining many exposures of the
spectrograph illuminated by a halogen lamp.

\placetable{tbl-targets}

The data were reduced using the FIGARO and KAPPA
packages\footnote{see \texttt{http://star-www.rl.ac.uk/}}. The
position-velocity diagrams were obtained via standard reduction
processes: i.e., dark-subtraction, flat-fielding, removal of bad
pixels, night sky subtraction, correcting for curvature in the
echelle spectra, wavelength calibration and correcting for telluric
absorption. Wavelength calibrations were made using OH airglow and
telluric absorption lines giving an uncertainty in the calibration
of less than $\pm$4~km\,s$^{-1}$. Before correcting for telluric
absorption, Brackett-series absorption features were removed from
the spectra of the standard stars by Lorentzian fitting.

\begin{figure}
\epsscale{0.5}
 \plotone{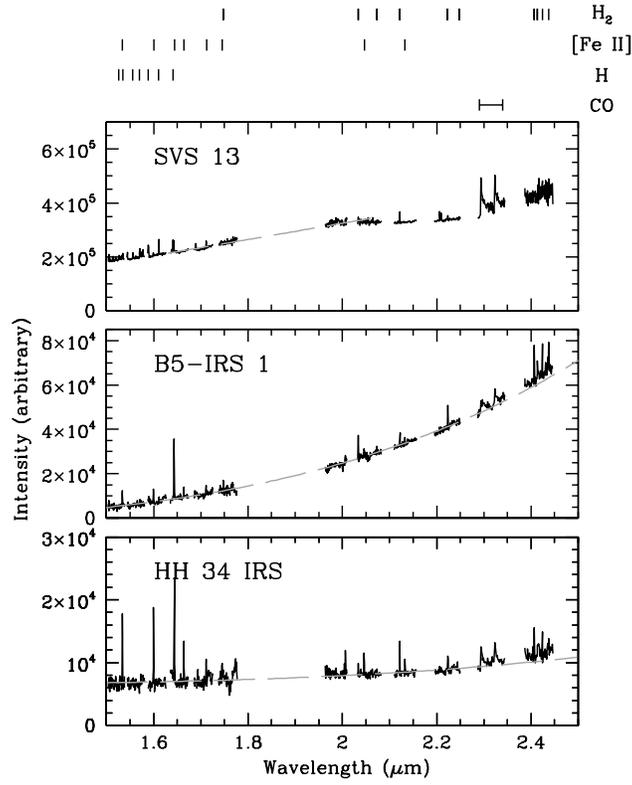} \caption{Spectra obtained
towards SVS 13, B5-IRS~1 and HH~34~IRS. The spectra are binned into
16 pixel bins, providing an actual spectral resolution $R$=2500.
Bright emission features are indicated at the top. The grey dashed
lines show the continuum level the spectra were scaled to (see
text).
}
  \label{whole_spec}
\end{figure}

The sky was not photometric, thus the relative flux between $H$- and
$K$-band spectra had to be carefully calibrated. We did this by
scaling one of the two spectra for each target so that a fit to the
continuum in the two bands could be made using a single linear or
polynomial function. For SVS 13, we were not able to fit the entire
continuum with a single linear/polynomial function due to the
presence of H$_2$O emission bands (see Carr, Tokunaga, \& Najita
2004). The fitting was thus carried out only for limited spectral
ranges near the boundary of the $H$- and $K$-bands
(1.63--1.78/1.96-2.05~$\mu$m). For B5-IRS~1 and HH~34~IRS, the
entire continuum was used for the fitting.
The results are shown in Figure \ref{whole_spec}. Fitting errors
suggest uncertainties for the relative fluxes between $H$- and
$K$-band spectra to be 2, 4 and 6\,\% for SVS 13, B5-IRS~1 and
HH~34~IRS, respectively. The 2.25/1.65~$\mu$m continuum flux ratio
measured in SVS 13 is 1.58, in excellent agreement with that in the
0.8--2.5~$\mu$m spectrum obtained by Carr et al. (2004).

\placefigure{whole_spec}


\section{H$_2$ 1--0 S(1) and [Fe II] 1.644~$\mu$m Emission}

\begin{figure}
\epsscale{0.8} \plotone{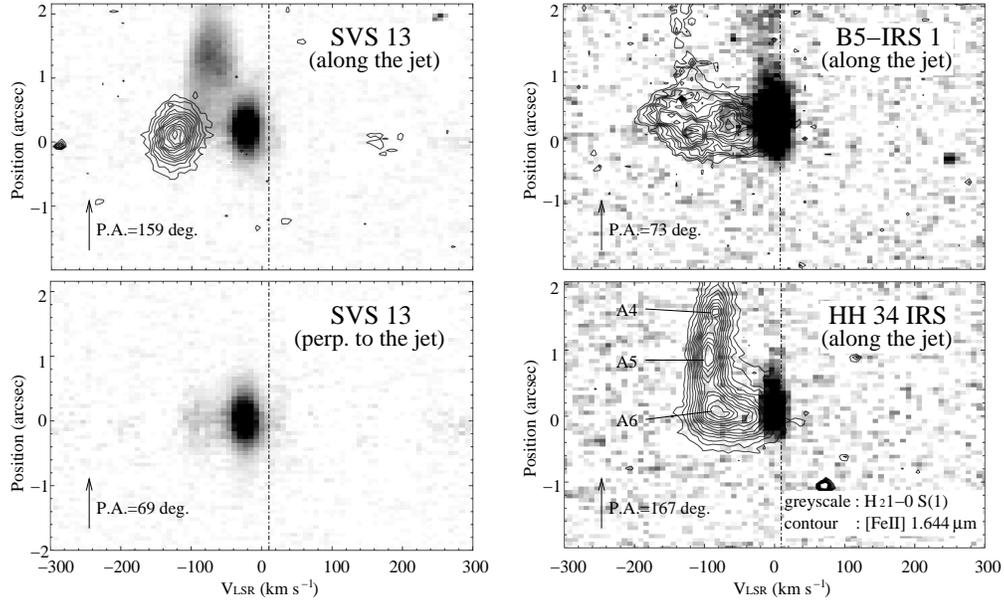} \caption{Continuum-subtracted
position velocity diagrams of H$_2$ 2.122~$\mu$m and [Fe II]
  1.644~$\mu$m emission. Greyscale and contours show the H$_2$ and
           [Fe II] emission, respectively. The contour spacing is 7\,\% of
the peak intensity. The dot-dashed line shows the LSR velocity of
the parent cloud. The zero spatial position of H$_2$ and [Fe II]
emission corresponds to the continuum centroid position adjacent to
each line. The continuum centroid positions at these wavelengths
coincide with each other to within an accuracy of 0.1\arcsec\ in all
objects (see Appendix B). The nomenclature for the knots in HH 34
are based on Reipurth et al. (2002). No contours are seen in the
lower-left figure since [Fe II] spectra were not obtained at this
slit position.}
  \label{pv}
\end{figure}

\begin{figure}
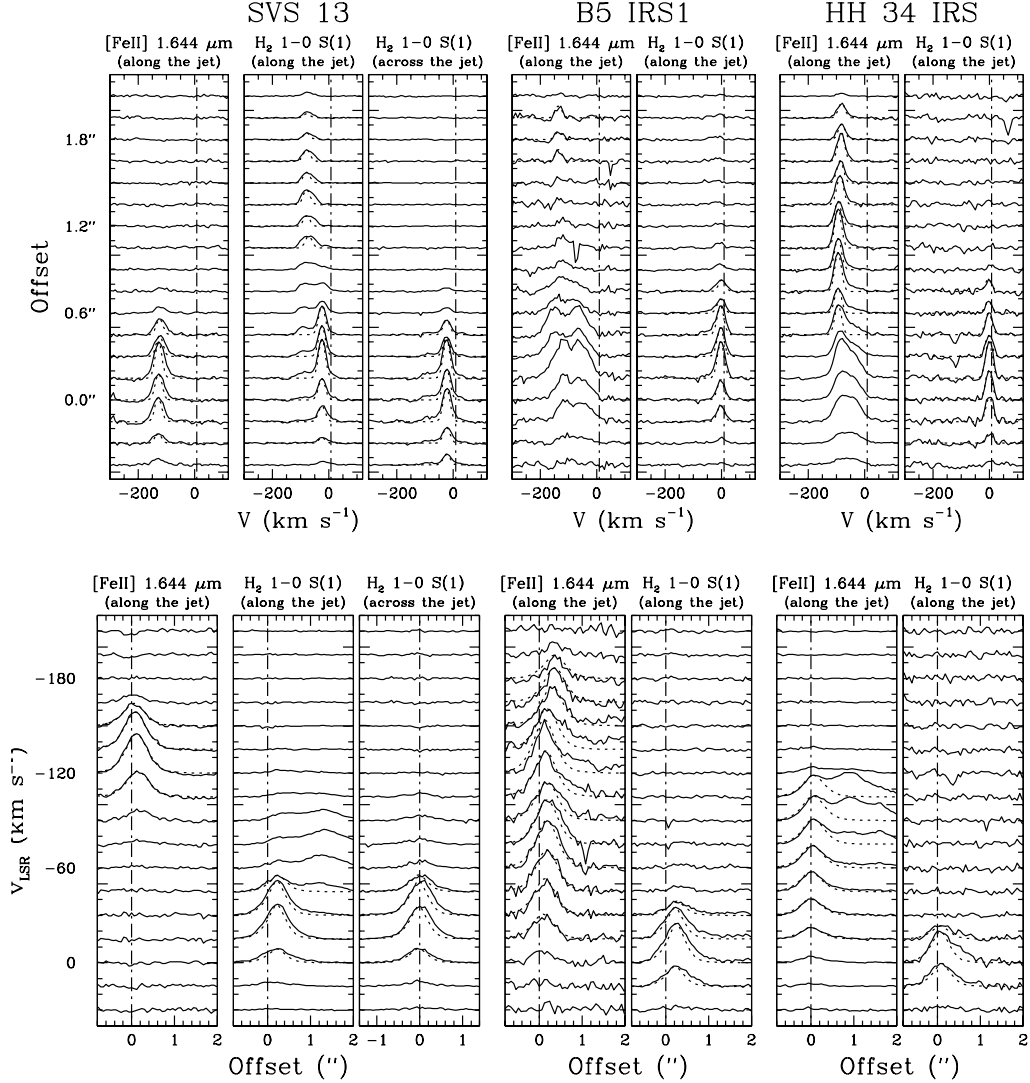

\epsscale{0.8} \plotone{f3a.epsi}  \hspace*{-0.5cm} \vspace{0.5cm}
\epsscale{0.82} \plotone{f3b.epsi} \caption{(\textit{top}) Line
profiles of [Fe II] 1.644~$\mu$m and H$_2$ 1--0 S(1) at individual
positional offsets from the star. Dotted curves show the
instrumental profile. The dot-dashed lines shows the systemic
velocity of the parent cloud. (\textit{bottom}) Intensity
distributions of the same lines at given velocities. Dotted curves
show the seeing profile. The dot-dashed lines shows the continuum
position of the star. }
  \label{profiles}
\end{figure}

Figure \ref{pv} shows the position-velocity (P-V) diagrams of the
H$_2$ 2.122~$\mu$m and [Fe II] 1.644~$\mu$m emission, the brightest
H$_2$ and [Fe II] lines. To show the kinematic and intensity
distribution in detail, we extract line profiles at different
positions and velocities and plot them in Figure \ref{profiles}. All
velocities quoted are in the local standard of rest (LSR) frame. In
the following subsections we describe the results for individual
objects.

\placefigure{pv} \placefigure{profiles}

\subsection{SVS 13}

Echelle spectroscopy by Davis et al. (2001, 2003) revealed the
following kinematic components in H$_2$ and [Fe II] emission: (1)
two velocity components in the H$_2$ emission peaking at --90 and
--20~km\,s$^{-1}$; and (2) two velocity components in the [Fe II]
emission at --130~km\,s$^{-1}$ and --20 to --50~km\,s$^{-1}$. The
authors measured typical spatial offsets for the H$_2$ emission of
0.4\arcsec\, and 0.2\arcsec\, from the star for the high- and
low-velocity components, respectively. Fabry-Perot imaging shows
that the H$_2$ emission at SVS 13 is associated with a collimated
small-scale jet (Davis et al. 2002).


Our results confirm the presence of three out of the four velocity
components identified by Davis et al. (2001, 2003). These are: the
two velocity components in H$_2$ and the high velocity component in
[Fe II].
In Figures \ref{pv} and \ref{profiles} the high-velocity component
(HVC) of H$_2$ emission is extended over a few arcsec, peaking at
1.3\arcsec\, with a FWHM velocity of $\sim$~70~km\,s$^{-1}$. This
component coincides with the small-scale H$_2$ jet seen in
Fabry-Perot images by Davis et al. (2002).
The centroid velocity of the H$_2$ HVC coincides with the high
velocity peak of HH 7--10, the extended blueshifted outflow
associated with SVS 13, observed in atomic and H$_2$ lines (Solf \&
B\"ohm 1987; Hartigan et al. 1987; Movsessian et al. 2000; Davis et
al. 2000, 2001). We thus conclude that this HVC component for H$_2$
is the base of the extended HH outflow.

The low velocity component (LVC) of the H$_2$ emission is only
marginally resolved, both spatially and spectroscopically. This
indicates that its spatial scale and velocity dispersion are
comparable to the seeing (0.6\arcsec) and spectral resolution
(28~km\,s$^{-1}$), respectively. The centroidal position of the
H$_2$ LVC is offset from the star by 0.2\arcsec--0.3\arcsec,
consistent with previous spectro-astrometric measurements (Davis et
al. 2001). Figure 3 shows that the H$_2$ emission across the jet
axis marginally extends to the north-east.

[Fe II] emission is spatially unresolved, with a centroid position
of $\sim$~0.1\arcsec\, offset from the continuum towards the jet.
The line exhibits a FWHM velocity of 36--38~km\,s$^{-1}$, slightly
larger than the spectral resolution of 28~km\,s$^{-1}$.
Interestingly, its centroid position decreases as the velocity
increases: 0.15\arcsec\, at $-$105~km\,s$^{-1}$ to 0.05\arcsec\, at
$-$150~km\,s$^{-1}$. This tendency contrasts with that normally
found for forbidden line emission in T Tauri stars, in which the
centroid position increases with velocity (see e.g., Takami et al.
2001; Whelan et al. 2004).


\subsection{B5-IRS~1}

High-resolution spectroscopy by Yu, Billawala, \& Bally (1999) and
Davis et al. (2001) measured an LSR velocity for the H$_2$
2.122~$\mu$m emission of $-$10 to 4~km\,s$^{-1}$, i.e.,
6--20~km\,s$^{-1}$ blueshifted from the systemic velocity of the
parent cloud. The H$_2$ emission did not show any spatial extension
at the driving source, while the presence of H$_2$ emission at HH
366 E5 was seen 20\arcsec\, away. Echelle spectroscopy by Davis et
al. (2003) revealed two components of [Fe II] emission associated
with the driving source, peaking at $-$140 and --70~km\,s$^{-1}$.
The high velocity emission at $-$140~km\,s$^{-1}$ was found to be
extended towards the jet.

The H$_2$ emission in Figures \ref{pv} and \ref{profiles} show two
components at similar low velocities -- a bright component peaking
at 0.2\arcsec--0.3\arcsec\, from the star, and a faint component
extending towards the jet over a few arcsec. The bright component
shows a centroidal LSR velocity of --2~km\,s$^{-1}$, about
$\sim$~10~km\,s$^{-1}$ blueshifted from the systemic velocity of the
parent molecular cloud. This component is only marginally resolved
both spatially and spectroscopically (with seeing at 0.6\arcsec\,
and spectral resolution of 28~km\,s$^{-1}$). The centroid velocity
of the H$_2$ emission increases with distance: --7 and --20 km
s$^{-1}$ at 1\arcsec\, and 2\arcsec\, from the star, respectively.

Similarly, the [Fe II] emission also consists of two components: a
bright component within an arcsec of the star and an extended
component towards HH 366E. The bright component is not spatially
resolved, and offset 0.4\arcsec\, towards the jet. This component
shows a wide velocity range (0 to $-$200~km\,s$^{-1}$) with the
position gradually increasing with velocity. Figure \ref{pv} shows
three peaks for this component, appearing at positions:
($-$75~km\,s$^{-1}$, 0.2\arcsec), ($-$115~km\,s$^{-1}$, 0.1\arcsec),
and ($-$160~km\,s$^{-1}$, 0.4\arcsec). The faint extended component
is at a velocity of $\sim -$140~km\,s$^{-1}$, and is not
spectroscopically resolved. The velocity differs remarkably from HH
366E ($\sim$--20~km\,s$^{-1}$ --- Bally, Devine, \& Alten 1996),
however, its low velocity dispersion relative to the centroid
velocity suggests that the emission is associated with a collimated
jet (see Eisl\"offel et al. 2000a).


\subsection{HH~34~IRS}
Echelle spectroscopy by Davis et al. (2001) showed that H$_2$
emission at HH 34 IR has a velocity of 1~km\,s$^{-1}$, which is
$\sim$~10~km\,s$^{-1}$ blueshifted from the systemic velocity of the
parent cloud. Davis et al. (2003) showed that [Fe II] emission at
the driving source peaks at --95~($\pm$10)~km\,s$^{-1}$. The line
profile they observed shows an extended wing towards the systemic
velocity of the parent cloud.

In our spectra, the H$_2$ emission shows a LSR velocity range of
$-$20 to 20~km\,s$^{-1}$. It is not spectroscopically resolved,
indicating that the velocity dispersion is smaller than
28~km\,s$^{-1}$. The peak H$_2$ emission is spatially offset by
$\sim$~0.1\arcsec\, and shows faint extension toward the jet up to
1\arcsec\, from the star. The HVC is not seen in H$_2$ towards HH
34, in contrast to SVS 13 and analogous to B5-IRS~1.

The [Fe II] emission in Figures 2 and 3 shows a broad component near
the base of the jet while the jet itself is narrower in the diagram.
Such a result is consistent with previous echelle spectroscopy by
Davis et al. (2003). In contrast to B5-IRS~1, these two components
are similarly bright. The former shows a single peak at the velocity
of the jet ($-$90 to $-$100~km\,s$^{-1}$
--- Davis et al. 2003), while the latter shows three peaks at
($-$90~km\,s$^{-1}$, 0.1\arcsec), ($-$100~km\,s$^{-1}$, 0.9\arcsec),
and ($-$90~km\,s$^{-1}$, 1.5\arcsec). These positions coincide with
the knots in the jet seen in $HST$ images, which peak at
0.0\arcsec\, (A6), 0.5\arcsec\, (A5), and 1.2\arcsec\, (A4) from the
driving source (Reipurth et al. 2002). Slight differences in
positions between these two observations are attributed to their
proper motions, which are 0.11/0.13/0.02 arcsec/yr for A4/A5/A6,
respectively (Reipurth et al. 2002).

\section{H$_2$ and [Fe II] Lines and Inferred Extinction, Temperature and Density}

\begin{figure}
\epsscale{1.0} \plotone{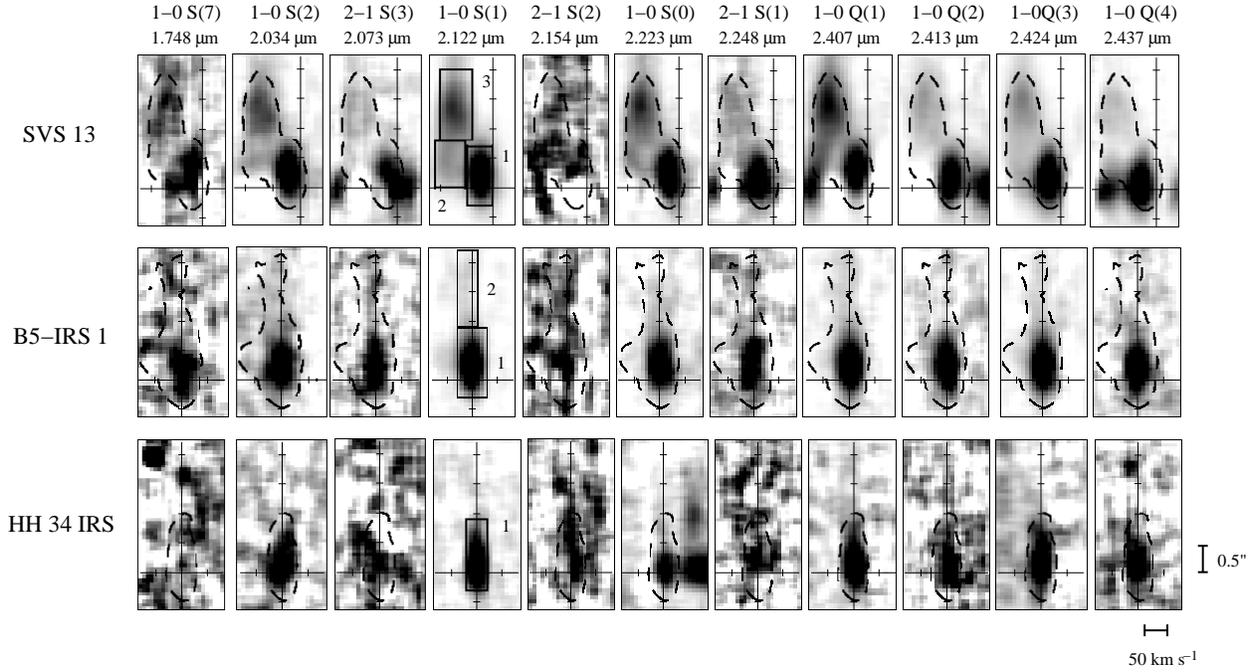} \caption{Continuum-subtracted
position-velocity diagrams of H$_2$ lines. The greyscale is
  arbitrarily adjusted for each panel to enhance the faint emission.
  Dashed curves show a single contour of 1--0 S(1) emission, corresponding
  to 15, 5, 10\,\% of the peak intensity for SVS 13, B5-IRS~1 and HH 34
  IRS, respectively. In the diagram for 1--0 S(1) emission we indicate the
  regions where the line intensities given in Table 2 are measured. Note that
  some panels for SVS 13 also contain emission/absorption features at
  the star. We also see bright emission at positive velocities relative to the 1--0 S(0) line for HH~34~IRS. This
  is due to either a ghost or an unidentified line associated with the object.}
\label{PV_H2}
\end{figure}

\begin{figure}
\epsscale{1.0} \plotone{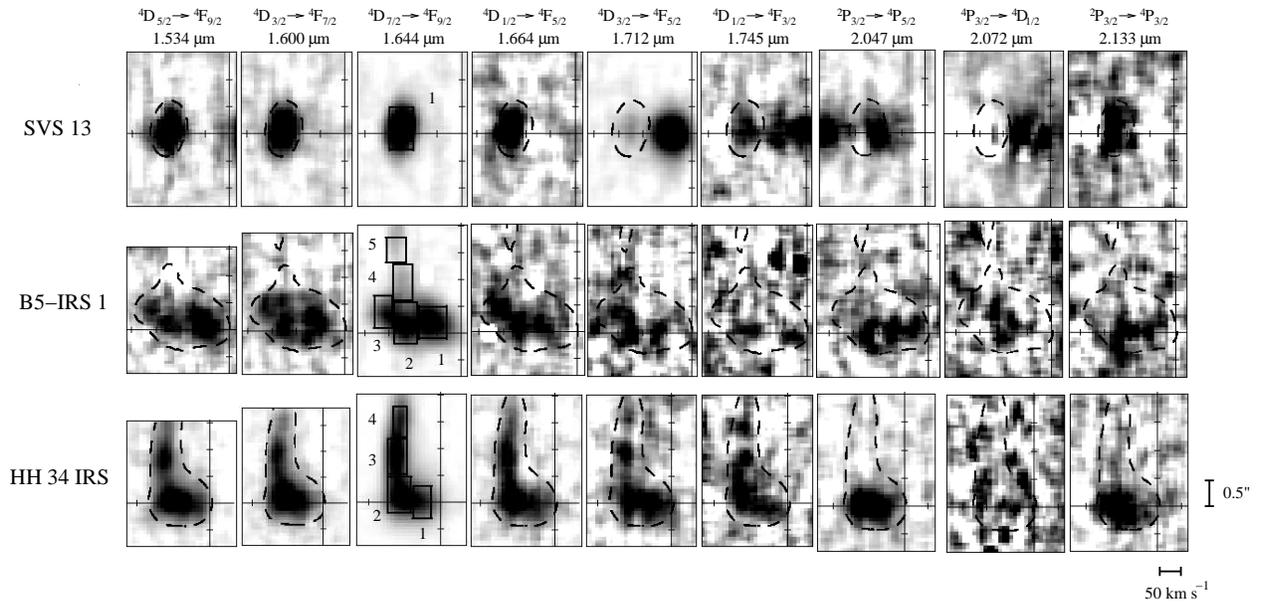} \caption{ The same as Figure
\ref{PV_H2} but for [Fe II] lines. Dashed curves show a single
contour of 1.644~$\mu$m emission, corresponding to 20\,\% of the
peak intensity. In the diagram for 1.644~$\mu$m emission we indicate
regions where the line intensities given in Table 2 are measured.}
  \label{PV_FeII}
\end{figure}

The spectra presented here contain a number of H$_2$ and [Fe II]
lines in addition to the bright H$_2$ 1--0 S(1) 2.122~$\mu$m and [Fe
II] 1.644~$\mu$m lines. These include H$_2$ emission arising from
$v=1-3$ and $J=1-11$, and [Fe II] emission with upper state energies
$E/k=1-3\times10^4$~K. It is noteworthy that we detect a few [Fe II]
$^2P \rightarrow {^4P}$ lines in the $K$-band, which is presumably
their first detection in YSOs. Figures \ref{PV_H2} and \ref{PV_FeII}
show PV diagrams of bright H$_2$ and [Fe II] lines, respectively.
Although the signal-to-noise ratios of some lines are not high
enough to investigate their kinematic structure, the H$_2$ lines
appear similar to those of H$_2$ 1--0 S(1). Regarding [Fe II]
emission, the $^4D \rightarrow {^4F}$ lines all show similar
kinematic structure, while $^2P \rightarrow {^4P}$ lines appear
spatially compact at the driving source.

\placefigure{PV_H2} \placefigure{PV_FeII}

The relative line fluxes in the regions shown in Figures \ref{PV_H2}
and \ref{PV_FeII} were measured and tabulated in Tables
\ref{tbl-H2ratios} and \ref{tbl-FeIIratios}. These regions are
labeled as Regions 1 to $N$ ($N$=1--5) for each object and line
(H$_2$/[Fe II]). The measured flux ratios are used in the following
sections to investigate the physical conditions of, and extinction,
to the H$_2$ and [Fe II] emission line regions.

\placetable{tbl-H2ratios} \placetable{tbl-FeIIratios}


\subsection{Extinction}
If two emission lines share the same upper energy level, their flux
ratio depends only on the extinction and physical constants. The
measured flux ratio of these lines can thus be used to determine the
extinction towards the emission line regions. Such pairs in our
spectra include:  H$_2$ 1--0 S(1) 2.122~$\mu$m and 1--0 Q(3)
2.424~$\mu$m; 1--0 S(0) 2.223~$\mu$m and 1--0 Q(2) 2.413~$\mu$m;
1--0 S(2) 2.034~$\mu$m and 1--0 Q(4) 2.437~$\mu$m; [Fe II]
$a^4D_{3/2} \rightarrow a^4F_{7/2}$ 1.600~$\mu$m and $a^4D_{3/2}
\rightarrow a^4F_{5/2}$ 1.712~$\mu$m; $a^4D_{1/2} \rightarrow
a^4F_{5/2}$ 1.664~$\mu$m and $a^4D_{1/2} \rightarrow a^4F_{3/2}$
1.745~$\mu$m. Tables 4 and 5 show extinction obtained using these
lines, adopting transition probabilities for H$_2$ and Fe$^+$ by
Turner, Kirby-Docken \& Dalgarno (1977) and Nussbaumer \& Storey
(1988), respectively. For an extinction law we adopt $A_\lambda =
A_V \times \lambda ^{-1.6}$, based on $A_H/A_V=0.175$ and
$A_K/A_V=0.112$ obtained for diffuse interstellar clouds (Rieke \&
Lebofsky 1985).

\placetable{tbl-H2phys} \placetable{tbl-FeIIphys}

Different pairs of emission lines result in markedly different
values for the extinction. For instance, in H$_2$ Region 1 of SVS
13, the extinction measured from all three ratios significantly
differ, $A_V$ ranging from $<$1 to $\sim$15. In contrast, values
derived from the S(0)/Q(2) and S(1)/Q(3) ratios are the same as each
other in Region 1 of B5 IRS 1 within an uncertainty of 20\%.
Concerning [Fe II] emission, extinction obtained from the
1.712/1.600~$\mu$m ratio is systematically lower than those obtained
from the 1.745/1.664~$\mu$m ratio. The discrepancies in $A_V$ are
greater than a factor of two in some cases. This might be
interpreted as the assumed power law ($A_V \times \lambda ^{-1.6}$)
being invalid. However, a power law index of --1 or --2 instead of
--1.6 changes the discrepancies by only 2--3 \%, due to a similarity
of wavelengths between line pairs. Therefore, this cannot explain
the discrepancies in $A_V$ described above.

Systematic discrepancies of extinction obtained from [Fe II]
1.712/1.600~$\mu$m and 1.745/1.664~$\mu$m ratios may be due to
errors in adopted transition probabilities. Possible errors for [Fe
II] transitions are also reported by Bautista \& Pradhan (1998), who
studied optical [Fe II] spectra in the Orion Nebula. Regarding H$_2$
emission, Q-branch lines are in poorly transmitting atmospheric
windows, and calibration errors by telluric absorption may also
contribute to the apparently different extinctions described above.

Such uncertainties in the extinction will affect the determination
of gas temperatures and densities in the following subsections. For
the H$_2$ emission line regions we adopt the extinction obtained
from the S(1)/Q(3) ratio for the following reasons: (1) consistency,
as this ratio is often used in the literature (e.g., Gredel 1994;
Eisl\"offel, Smith, \& Davis 2000b); (2) when this ratio is obtained
at the highest S/N ratio agrees with those measurements using the
S(0)/Q(2) and/or S(2)/Q(4) ratios; and (3) both S(1) and Q(3) lines
are in relatively clean parts of the atmospheric window. For [Fe II]
emission we regard any extinction obtained from the
1.712/1.600~$\mu$m ratio as upper limits.

\subsection{Excitation, Temperature and Density in H$_2$ Emission Line Regions}
\begin{figure}
\epsscale{0.8} \plotone{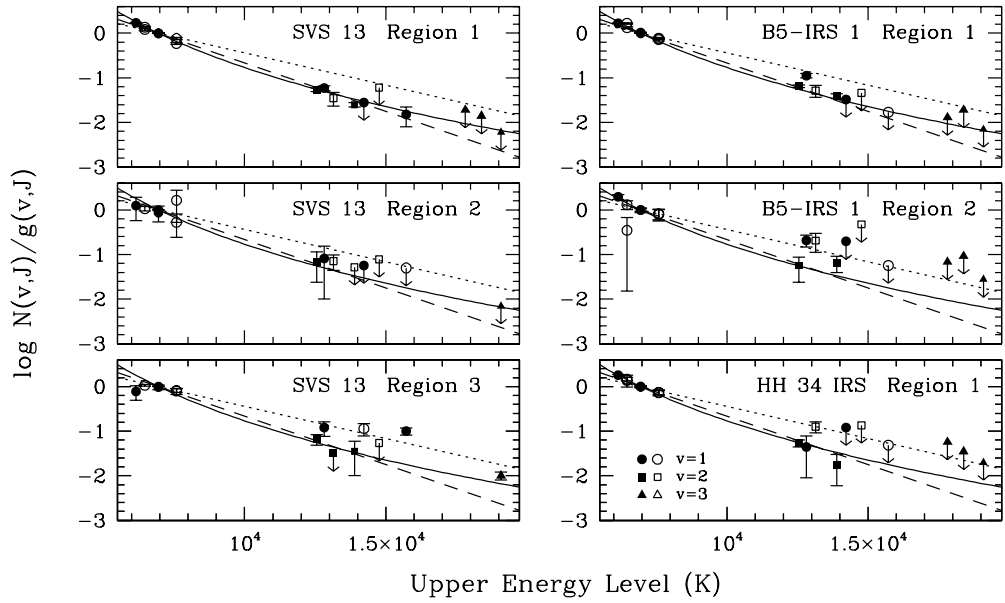}
  \caption{Population diagram of H$_2$ emission for regions shown in Figure 4.
           Circles, squares and triangles indicate the populations at $v=1,
           2, 3$ respectively. Filled and open marks indicate those for ortho
           and para H$_2$, respectively. Dashed and dotted lines indicate
           an LTE population at $T$=2000 and 3000 K, respectively. Solid
           curves show the populations for a dissociative J-shock modelled by Brand et al. (1988).}
  \label{H2pop}
\end{figure}

Figure \ref{H2pop} shows a population diagram of ro-vibrationally
excited H$_2$ obtained from our spectra. The figure shows that
molecular hydrogen in all regions is thermally excited at a
temperature of $2-3\times10^3$~K. In particular, the temperature
obtained from the 2--1 S(1)/1--0 S(1) flux ratio is consistently
stable at $2\times10^3$~K for all regions and objects. These trends
are often observed in shocks associated with extended outflows
(e.g., Burton et al. 1989; Gredel 1994; Eisl\"offel et al. 2000b).
Thermal excitation of infrared H$_2$ emission could also be attained
via UV/X-ray heating (e.g., Burton, Hollenbach \& Tielens 1990;
Maloney, Hollenbach, \& Tielens 1996; Tine et al. 1997) or ambipolar
diffusion (Safier 1993). However, these models predict that line
flux ratios strongly depend on the radiation field and gas density,
or distance from the driving source. We thus conclude that shock
excitation is a robust mechanism for the H$_2$ emission.

We can fit the level populations in Figure 6 using a single
temperature line or curve for most of the regions and objects. In
contrast, in Region 3 of SVS 13, the vibrational temperature at
$v=1$ and $J=0-3$ shows a temperature of $2\times10^3$~K, lower than
the rotational temperature at $v=1$ ($3\times10^3$~K). This
indicates that the gas in this region has not attained LTE.
Performing comparisons with population diagrams modelled by
Gianninni et al. (2002), we estimate a H$_2$ number density of $\sim
10^7$~cm$^{-3}$ for Region 3 of SVS 13, and $\sim 10^8$~cm$^{-3}$ or
greater for the other regions, assuming that thermal excitation is
dominated by H$_2$--H$_2$ collisions. Collisional rate coefficients
for H--H$_2$ collisions are $\sim$~100 times larger than
H$_2$--H$_2$ collision for temperatures of $2-3\times10^3$~K (see
e.g., Burton et al. 1990) thus the same results are also explained
by hydrogen atomic number densities of $\sim~10^5$~cm$^{-3}$ for
Region 3 of SVS 13, and $\sim 10^6$~cm$^{-3}$ or greater for the
other regions. In either case, our results for SVS 13 suggest that
the gas density decreases as one moves downstream of the shock:
i.e., Region 2 to Region 3. This trend is also observed in HH
outflows (e.g., Bacciotti \& Eis\"offel et al. 1999) and also
so-called ``micro-jets'' in T Tauri stars (e.g., Lavalley-Fouquet,
Cabrit, \& Dougados 2000; Bacciotti et al. 2000; Woitas et al. 2002;
Dougados, Cabrit, \& Lavalley-Fouquet 2002).

\placefigure{H2pop}

H$_2$ population diagrams also allow us to discriminate between
types of shock, e.g. C- and J-type shocks. Models predict that for
energy levels $E/k=0.5-2\times10^4$~K the level populations indicate
a rather constant temperature for planar C-shocks while higher
temperatures are required to excite the higher energy levels in
planar J-shocks (see Eisl\"offel et al. 2000b, and references
therein). To investigate this in detail, we show in Figure
\ref{H2pop} the J-shock model prediction which empirically explains
the H$_2$ populations observed in Orion KL (Brand et al. 1988) and
supernova remnant IC 443 (Richter et al. 1995). The figure shows
that the population in Region 3 of SVS 13 cannot be attributed to
either a planar J-shock or an LTE model with a single temperature,
representative of a planar C-shock. J/C-shock models with non-LTE
populations and/or more complex geometries may be required to
interpret the observations. For the other regions and objects, LTE
seems to have been attained, although higher S/N data are required
to discriminate between C- and J-shocks.

Fernandes \& Brand (1995) observed a number of near-IR H$_2$
emission in HH 7, and argue the presence of UV-pumped H$_2$ emission
based on their results. Although this excitation mechanism can
provide non-LTE H$_2$ populations, it cannot explain the non-LTE
population observed in Region 3 of SVS 13. In Figure \ref{H2pop},
the population at $v$=1 and $J$=1--11 indicates a rotational
temperature of $\sim$3$\times$10$^3$ K in this region, while that at
$v$=1--2 and low-$J$ levels indicates a vibrational temperature of
$\sim$2$\times$10$^3$ K, lower than the other. This trend is
opposite as observed in UV-pumped H$_2$ regions, in which rotational
temperatures are lower than vibrational temperatures (see e.g.,
Hasegawa et al.; Tanaka et al. 1989).

\subsection{Electron Density and Temperature in [Fe II] Emission Line Regions}
To investigate electron densities and temperatures of the [Fe II]
emission line regions, we have developed a non-LTE model which
considers the first 16 fine structure levels.
Transition probabilities are taken from Nussbaumer \& Storey (1988),
the energy levels are from the NIST Atomic Spectra
Database\footnote{\texttt{http://physics.nist.gov/cgi-bin/AtData/main\_asd}},
and the rate coefficients for electron collisions are from Zhang \&
Pradhan (1995). All radiative transitions among these levels are
quadrupole or magnetic dipole type, thus we assume optically thin
conditions and compute level populations under statistical
equilibrium. This model is essentially the same as that developed by
Zhang \& Pradhan (1995), Nisini et al. (2002) and Pesenti et al.
(2003), although we provide the following two new line flux ratios:
$I_{1.664\,\mu m}/I_{1.644\,\mu m}$, which is not sensitive to
extinction, and, $I_{1.749\,\mu m}/I_{1.644\,\mu m}$, which is
highly sensitive to the electron temperature as well as the electron
density.

\begin{figure}
\epsscale{0.9} \plotone{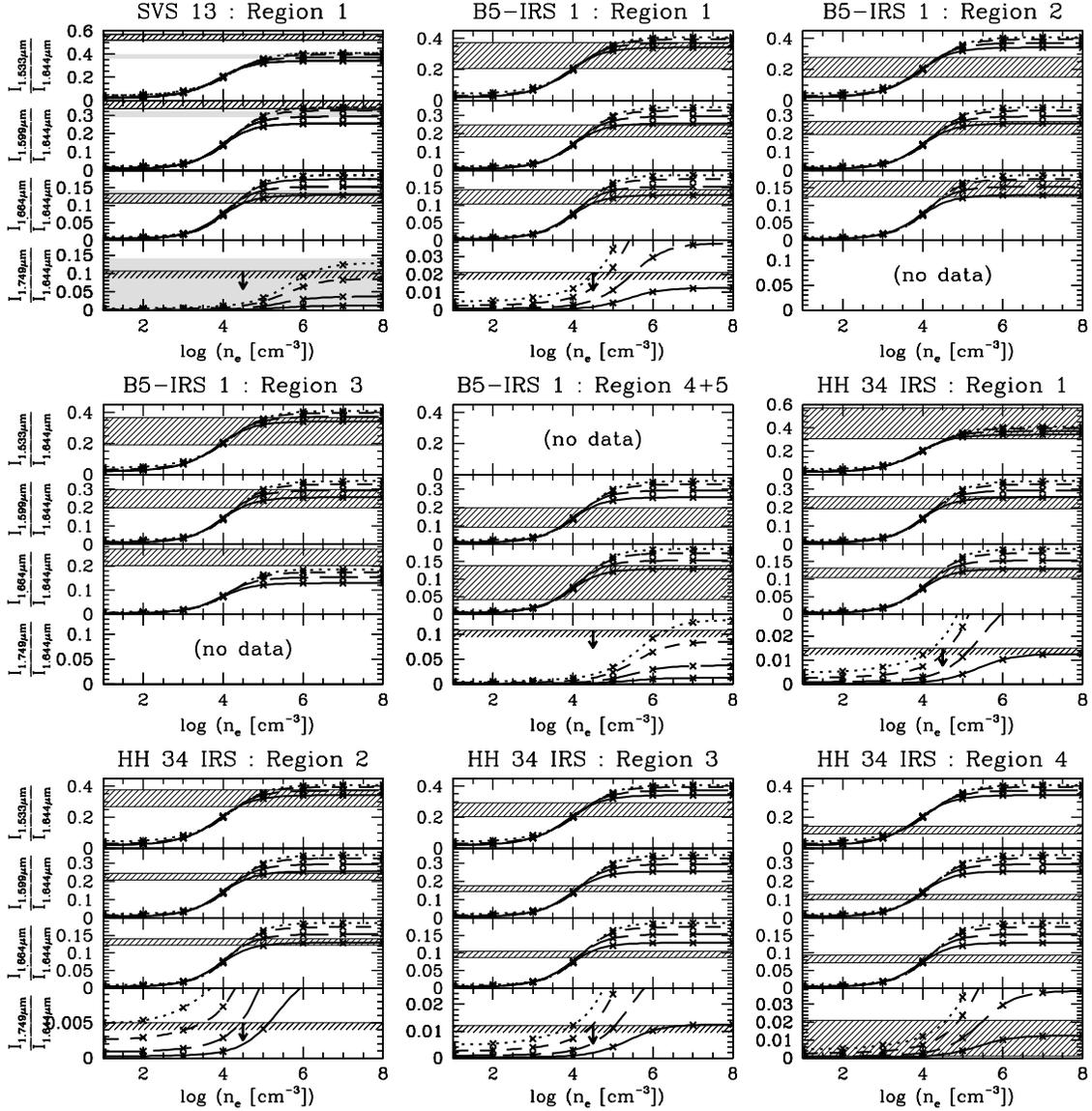} \caption{Modeled and observed [Fe
II] line flux ratios. Modeled values are shown as solid,
long-dashed, short dashed and dotted lines for
$T_e$=0.3/0.5/1/2$\times$10$^4$ K, respectively.
The hatched regions indicate the observed flux ratios and their
uncertainty. For SVS 13, hatched and grey regions
show the ratios for $A_V=20$ and 0, respectively. Upper limits are
used for extinction towards the other objects (see Table 5) which is
propagated into the observed ratios shown here.}
  \label{FeIIratios}
\end{figure}

Figure \ref{FeIIratios} shows the modelled and observed line flux
ratios. For SVS 13 we assume either no extinction ($A_V=0$) or
$A_V=20$, obtained using the $I_{1.257\,\mu m}/I_{1.644\,\mu m}$
ratio by Gredel (1996). The $I_{1.533\,\mu m}/I_{1.644\,\mu m}$
ratio is the least sensitive to the electron temperature, and these
two lines are brighter than the other $^4D \rightarrow {^4F}$
transitions such as the 1.600 and 1.664~$\mu$m lines (see Table 3).
This ratio is thus often used to derive an electron density (see
e.g., Hamann et al. 1994; Itoh et al. 2000; Pesenti et al. 2003).
However, Figure 7 shows that a combination of $I_{1.600\,\mu
m}/I_{1.644\,\mu m}$ and $I_{1.664\,\mu m}/I_{1.644\,\mu m}$ ratios
provide electron densities with better accuracies in some cases
since these are not much affected by the uncertainty in extinction
described in \S 4.1.

Figure \ref{FeIIratios} shows that the electron density of [Fe II]
emission line regions $n_e$$\sim$10$^4$ cm$^{-3}$ or greater in all
the objects and regions we observe. Such a density is similar to [Fe
II] emission line regions associated with T Tauri stars (Hamann et
al. 1994). In B5-IRS~1 and HH~34~IRS, $I_{1.664\,\mu
m}/I_{1.644\,\mu m}$ and $I_{1.600\,\mu m}/I_{1.644\,\mu m}$ (and
also the $I_{1.533\,\mu m}/I_{1.644\,\mu m}$ ratios in HH~34~IRS)
are lower at the extended part than the base, indicating that the
electron density decreases downstream of the jet. As previously
stated such a trend is also observed in previous observations of
extended HH outflows (e.g., Bacciotti \& Eis\"offel et al. 1999) and
also ``micro-jets'' in T Tauri stars (e.g., Lavalley-Fouquet et al.
2000; Bacciotti et al. 2000; Woitas et al. 2002; Dougados et al.
2002).

\placefigure{FeIIratios}

As shown in Figure \ref{FeIIratios}, the $I_{1.533 \mu m}/I_{1.644
\mu m}$, $I_{1.600\,\mu m}/I_{1.644\,\mu m}$, $I_{1.664\,\mu
m}/I_{1.644\,\mu m}$ ratios are not sensitive to electron
temperature due to their upper energy levels being similar to each
other. The lines with significantly different upper energy levels,
such as $^4$$P$$\rightarrow$$^4$$D$ 1.749~$\mu$m, are required with
better accuracy to determine the electron temperature. The
temperature inferred by the $I_{1.749\,\mu m}/I_{1.644\,\mu m}$
ratio appears to be similar to that of forbidden emission line
regions associated with T Tauri stars ($\sim$10$^4$ K, e.g., Hamann
1994; Dougados et al. 2002) and extended HH objects ($\sim$10$^4$ K,
e.g., Bacciotti \& Eis\"offel 1999), although higher S/N ratios are
required to determine this parameter.

\subsection{[Fe II] $^2P \rightarrow {^4P}$ lines as a new probe for jets}
\begin{figure}
\epsscale{0.5} \plotone{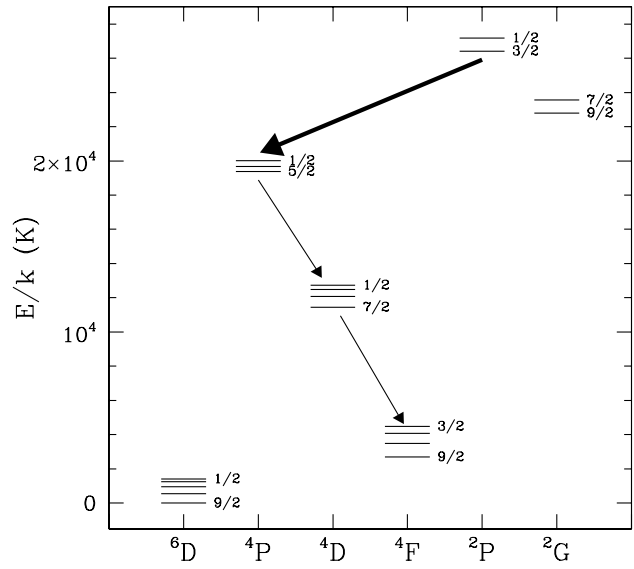}
  \caption{Lowest 20 levels of Fe$^+$. Arrows show transitions we detect
above the 3$\sigma$ level in at least one of our targets. The thick
arrow shows the $^2P \rightarrow {^4P}$ transitions.}
  \label{FeII_diagram}
\end{figure}

 Figure \ref{FeII_diagram} shows the energy level diagram of Fe$^+$
highlighting the transitions detected in this work. In addition to
transitions at the lowest 16 levels, which are often observed and
modelled, our spectra exhibit several $^2P \rightarrow {^4P}$ lines
between 2.007 and 2.244~$\mu$m. The upper level energies of these
transitions are $\sim~2.6\times10^4$~K, more than twice as large as
the 1.644~$\mu$m line ($1.1\times10^4$~K). Due to longer
wavelengths, these lines suffer from less extinction than the bright
[Fe II] lines in the $H$-band: A$_\lambda$ is 30--40 \% smaller than
for the 1.644~$\mu$m line.

\placefigure{FeII_diagram}

\begin{figure}
\epsscale{0.8} \plotone{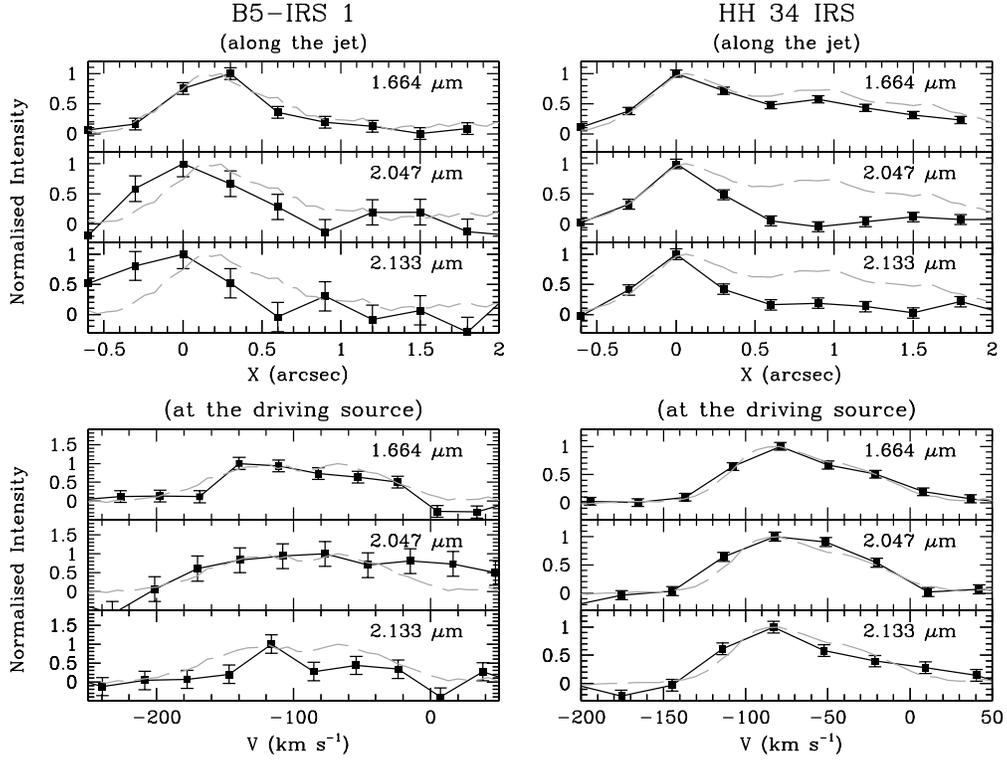}
  \caption{Spatial distribution (\textit{top}) and line profiles
  (\textit{bottom}) of [Fe II] 1.664/2.047/2.133~$\mu$m lines in B5-IRS~1
  (left) and HH~34~IRS (right). Those of [Fe II] 1.644~$\mu$m are also
  shown in each plot by grey solid lines. The spatial distribution is
  measured along the jet, by binning the intensity from $-$155 to
  $-$100~km\,s$^{-1}$ in B5-IRS~1 and $-$120 to $-$60~km\,s$^{-1}$ in HH
  34 IRS. The line profiles are extracted at the driving source, i.e.,
  $-$0.2\arcsec\, to 0.7\arcsec\, at B5-IRS~1 and $-$0.3\arcsec\, to
  0.5\arcsec\, at HH~34~IRS.}
  \label{FeII_dist}
\end{figure}

Figure \ref{FeII_dist} shows the spatial distributions and line
profiles of four [Fe II] emission lines, including two $^2P
\rightarrow {^4P}$ lines at 2.047 and 2.133~$\mu$m. In HH~34~IRS
that the 2.047 and 2.133~$\mu$m emission is concentrated at the base
of the jet is in contrast to the 1.664 and 1.644~$\mu$m emission. At
the driving source, the 2.047 and 2.133~$\mu$m emission have
profiles similar to 1.644~$\mu$m emission. These trends are not
clear in B5-IRS~1, although higher S/N ratios are required to
investigate them in detail.

\placefigure{FeII_dist}

The compact spatial distribution, and also the line profiles, of the
$^2P \rightarrow {^4P}$ lines at HH~34~IRS indicates that these
lines could be powerful probes for studying the physical conditions
of the flow accelerating region. HI recombination lines such as
Br$\gamma$ have been used for over a decade as investigative tools
of the jet/wind launching region (e.g., Natta, Giovanardi, \& Palla
1988; Nisini, Antonucci, \& Giannini 2004). However, the emission
could originate from magnetospheric accretion columns even in Class
I protostars as well as T Tauri stars (see e.g., Muzerolle et al.
1998; Davis et al. 2001). In contrast to this, but similar to the
FEL regions in Class I protostars and T Tauri stars, the [Fe II]
2.047 and 2.133~$\mu$m lines show only blueshifted components
indicating that these are associated with outflowing gas. The
absence of redshifted emission is attributed to obscuration of the
counter flow by a disk or flattened envelope (see e.g., Appenzeller
et al. 1984; Edwards et al. 1987).

Excitation conditions of the $^2P \rightarrow {^4P}$ transitions are
not clear due to large uncertainties in the collisional rate
coefficients to the $^2P$ level (see Zhang \& Pradhan 1995; Verner
et al. 1999). The spatial distribution observed at HH~34~IRS is
either due to high temperature or density at the jet launching
region. In addition, Verner et al. (2000) show that the level
population of the $^2P$ states is also affected by UV pumping, which
could be efficient within an AU of YSOs. Accurate determination of
collisional rate coefficients is desired to investigate the physical
conditions of the jet/wind launching regions in detail.

\section{Discussion}
\subsection{Comparison with T Tauri Stars --- Flow Geometry and Heating}
Jets/winds associated with T Tauri stars exhibit forbidden line
emission seen within 10--100 AU of the star. The lines often exhibit
two velocity components at high ($-$50 to $-$200~km\,s$^{-1}$) and
low velocities ($-$5 to $-$20~km\,s$^{-1}$
--- see Eisl\"offel et al. 2000a for a review). High-resolution
observations with the {\it Hubble Space Telescope} and ground-based
adaptive optics have confirmed that the high-velocity component
(hereafter HVC) is associated with a collimated jet (see e.g.,
Bacciotti et al. 2000; Woitas et al. 2002; Pyo et al. 2003).
Emission line diagnostics to date suggest that this component is
heated by shocks in internal working surfaces (see e.g.,
Lavalley-Fouquet et al. 2000; Takami et al. 2002; Dougados et al.
2002). In contrast to the HVC, the nature of the low velocity
component (hereafter LVC) is still debated, although it appears to
be associated with a compact ($<$~50~AU) and wide-angled component
of the flow (e.g., Bacciotti et al. 2000; Pyo et al. 2003). In
addition to T Tauri stars, such a two-component kinematic structure
to forbidden line emission is also observed at L1551-IRS 5, which is
in transition from a Class I protostar to a T Tauri star (Pyo et al.
2002, 2005).

Davis et al. (2001) has shown that the H$_2$ 2.122~$\mu$m emission
at HH driving sources has a similar kinematic structure to FEL
regions in T Tauri stars. The similarities include: (1) both MHEL
and FEL regions show blueshifted components at high (--50 to
--150~km\,s$^{-1}$) and low (--5 to --20~km\,s$^{-1}$) velocities;
(2) LVCs are more common than HVCs (see Hartigan et al. 1995 for
FELs); (3) the HVC is spatially further offset from the exciting
source than the LVC. Davis et al. (2003) showed that both H$_2$ and
[Fe II] 1.644~$\mu$m emission tend to be associated with each HH
driving source. In their spectra the [Fe II] is associated with
higher velocity gas than the H$_2$, and peaks further away from the
driving source in each system. This can be explained if the [Fe II]
is more closely associated with HH-type shocks in the inner, on-axis
jet regions, while the H$_2$ is excited along the boundary between
the jet and the near-stationary, dense ambient medium that envelopes
the protostar.

As shown in \S 3, outflowing gas close to SVS 13, B5-IRS~1 and
HH~34~IRS also show two blueshifted components i.e., a  HVC
associated with an extended collimated jet, and a LVC which is
spatially more compact. In contrast to T Tauri stars, the LVC in all
of these objects are observed in H$_2$ emission, and none of them
show a low velocity peak in [Fe II] emission. The observed line
ratios indicate a temperature for the H$_2$-LVC of
$2-3\times10^3$~K, significantly lower than the LVC observed in T
Tauri stars ($\sim~10^4$~K --- Hamann 1994). The different
temperatures arise from the fact that H$_2$ is dissociated at
T$>$4000 K, while a higher temperature is required to thermally
excite Fe$^+$ and thus produce near-IR [Fe II] emission.

In SVS 13, the HVC is observed in H$_2$ emission as well as the LVC.
This contrasts to the HVC in B5-IRS~1 and HH~34~IRS, which are only
observed in [Fe II]. The observed velocity of
$\sim~-70$~km\,s$^{-1}$ in the H$_2$ HVC is nearly the same as for
the forbidden line emission in the extended HH flow (Solf \& B\"ohm
1987; Hartigan, Raymond, \& Hartmann 1987; Movsessian et al. 2000).
This suggests that H$_2$ is excited by shocks in the internal
working surfaces, as observed in the HVC of T Tauri stars.

How does molecular hydrogen to survive the jets/winds associated
with Class I protostars? A few possible explanations come to mind.
Bacciotti \& Eisl\"offel (1999) have shown that the hydrogen
ionization fraction generally deceases along the HH jet, suggesting
the presence of a ``prompt ionization'' mechanism close to the
source. Indeed, the presence of Alfv\'{e}n wave heating or X-ray
heating have been suggested at the base of outflows associated with
classical T Tauri stars (e.g., Hartmann, Avrett, \& Edwards 1982;
Shang et al. 2002; Takami et al. 2003). Such ionization mechanisms
could be less efficient in these Class I protostars due to, e.g.,
weaker chromospheric activity, thereby allowing molecular hydrogen
to survive. Other explanations include: (1) the outflowing gas has a
higher gas density, providing a high cooling efficiency behind the
shocks; (2) magnetic cushioning occurs efficiently, leading to low
levels of excitation and dissociation (e.g., Draine, Roberge \&
Dalgarno 1983); and (3) H$_2$ is reformed quickly, in perhaps just
$\sim$10 yrs at a hydrogen number density $n_H$$>$10$^8$ cm$^{-3}$
(Davis et al. 2002).

SVS 13 also exhibits [Fe II] emission, which cannot be simply
explained by either a HVC associated with the jet, or an LVC. The
measured velocity of $\sim~-130$~km\,s$^{-1}$ is remarkably higher
than the H$_2$ HVC and also of that measured in HH 7--10 ($-40$ to
$-90$~km\,s$^{-1}$), while it is lower than that measured in HH 11
($-180$ to $-200$~km\,s$^{-1}$
--- Solf \& B\"ohm 1987; Hartigan et al. 1987; Movsessian et al.
2000).
%
The [Fe II] emission could originate from a higher velocity
``spine'' bracketed within the H$_2$ jet at a lower velocity, as
suggested for the extended jets (Davis et al. 2003). However, our
results in \S 3.1 suggest that the offset of the [Fe II] emission
from the star decreases as the velocity increases. This trend is the
opposite of what is generally seen in optical observations of
outflowing gas from T Tauri stars (see e.g., Takami et al. 2001;
Whelan et al. 2004). Based on preliminary proper motion studies,
Davis et al. (2006) postulate that the [Fe II] emission at the base
of SVS 13 might be excited in a stationary collimated shock in the
jet. More detailed spatial information is thus necessary to conclude
the nature of the [Fe II] emission associated with SVS 13.

\subsection{The Driving Mechanism of the Jet}
Understanding the mechanism which drives jets from YSOs is one of
the most important key issues for star formation. Radiation and
thermal pressure are not sufficient to explain their momentum (see
Reipurth \& Bally 2001), and there is growing evidence that these
flows are magneto-hydrodynamically driven. Possible mechanisms
include: magneto-centrifugal force (e.g., Blandford \& Payne 1982;
Shu et al. 2000; K\"onigl \& Pudritz 2000), magnetic pressure (e.g.,
Uchida \& Shibata 1985) and magnetic stress (e.g., Hayashi et al.
1996; Goodson et al. 1999).

Among these magneto-centrifugal wind models seem the most promising.
According to these models, magnetic and centrifugal forces act
together to launch the jet/wind along magnetic field lines, either
from a narrow region of the disk at $\le$~0.1~AU (``X-wind''
--- see Shu et al. 2000) or from 0.1 to a few tens of AU (``disk wind''
--- see, e.g., K\"onigl \& Pudritz 2000; Ferreira 1997). The magnetic
field lines act as solid wires up to the ``Alfv\'en surface'',
located at less than 20~AU from the disk. These accelerate the flow
particles outwards and upwards in a `bead-on-a-wire' fashion,
simultaneously removing angular momentum from the accretion disk.
Although present high-resolution facilities cannot resolve the
kinematics of the innermost region, the models seem to explain some
key observed properties, including the observed mass ejection per
mass accretion rates ($\sim$~0.1, see e.g., Calvet 1997; Richer et
al. 2000), and jet motion consistent with rotation around the axis
(Davis et al. 2000; Bacciotti et al. 2002; Coffey et al. 2004;
Woitas et al. 2005), and in the same sense as the circumstellar disk
or molecular envelope (Wiseman et al. 2001; Testi et al. 2002).

To date several authors have accepted the challenge of modelling the
forbidden line profiles of T Tauri stars based on
magneto-centrifugal wind models, using both X-wind (Shang, Shu, \&
Glassgold 1998) and disk-wind models (e.g., Cabrit et al. 1999;
Pesenti et al. 2003). However, these models do not reproduce well
the blueshifted peaks at high and low velocities (see e.g., Shang et
al. 1998; Cabrit et al. 1999) and the line profiles are very
different from the observed ones (see e.g., Garcia et al. 2001).
Pesenti et al. (2003) attempted in reproducing the two blueshifted
[Fe II] 1.644~$\mu$m emission peaks of L1551-IRS5 observed by Pyo et
al. (2002). However, the LVC modelled by these authors has a
remarkably smaller spatial scale and higher velocity than observed.
In addition, a model by the same authors does not well reproduce the
kinematic structure of [Fe II] emission in DG Tau. While the
observed line profile clearly shows a separated blueshifted peak at
a low velocity ($\sim$100 km s$^{-1}$) of the position-velocity
diagram (Pyo et al. 2003), this is marginal in their DG Tau model.

Unlike forbidden line emission from T Tauri stars, the [Fe II]
emission presented here for B5-IRS and HH~34~IRS does not show a low
velocity peak. Regarding this point, the P-V diagrams obtained for
these objects are similar to those of [S II] emission modelled by
Shang et al. (1998), and also the [Fe II] emission modelled for DG
Tau by Pesenti et al. (2003). Nevertheless, these two models have
significant differences. That of Shang et al. (1998), which is based
on the X-wind scenario, shows clear redshifted emission at the base
of the flow despite obscuration of the redshifted lobe by a
circumstellar disk. This redshifted component has a velocity range
similar to the blueshifted side in their model P-V diagrams due to a
wide opening angle and perhaps flow rotation in the accelerating
region. Pesenti et al. (2003) base their models on disk-winds and in
contrast to Shang et al. (1998), do not produce such redshifted
emission in their P-V diagrams, exhibiting a sharp intensity
`cut-off' near the zero velocity. In Figures 2 and 3, [Fe II]
emission in B5-IRS~1 and HH~34~IRS do not show the redshifted
emission predicted by Shang et al. (1998). Our results thus support
disk-wind models as a driving mechanism for jets, bringing us to the
same conclusion as that arrived at for jets in T Tauri stars
(Bacciotti et al. 2002; Anderson et al. 2003; Coffey et al. 2004;
Woitas et al. 2005).

\subsection{Origin of the Low Velocity Component}
As described in Sect 5.1, outflowing gas near T Tauri stars and
Class I protostars show two velocity components at high and low
velocities. While the origin of the HVC is well understood, that of
the LVC is still debated. Historically, a hollow-cone flow geometry
was first used to explain the two peaks (Appenzeller et al. 1984;
Edwards et al. 1987; Ouyed \& Pudritz 1993). This has been replaced
by the concept that the two components are the result of a
combination of different flows, e.g. stellar and disk-winds (e.g.,
Kwan \& Tademaru 1987, 1993); X-wind and disk-wind (Pyo et al.
2002); a reconnection wind and a disk-wind (Pyo et al. 2003). These
easily explain the different electron densities and ionizations
found for the two velocity components (e.g., Hamann 1994; Hartigan
et al. 1995; Hirth et al. 1997). In contrast, Calvet (1997) shows
that the forbidden line luminosities of the LVC are correlated with
those of the HVC, suggesting that these two components are
intimately related. This is corroborated by the {\it HST}
observations of Bacciotti et al. (2000), who reveal a continuous
bracketing of central high velocity gas within a lower velocity,
less collimated, broader flow, down to the lowest velocity scales in
DG Tau.

Adaptive optics observations by Pyo et al. (2003) show that the
velocity of the LVC in DG Tau increases with distance: $-$60 to
$-$120~km\,s$^{-1}$ at 0.2\arcsec\, to 0.7\arcsec\, from the star.
If this is due to acceleration, the spatial scale for this
acceleration is at least 10--100 times larger than that predicted by
models (see e.g., Shu et al. 1994; Ferreira et al. 1997). Pyo et al.
(2003) suggest that a significant part of the LVC is not the ejecta
of a magneto-hydrodynamically driven wind but gas entrained by the
HVC, i.e., a collimated jet.

The H$_2$-LVC in B5-IRS~1 exhibits faint extended emission, in which
the velocity increases with distance: --- $V_{LSR}$~=~--2 and
--20~km\,s$^{-1}$ at the driving source and 2\arcsec\, away,
respectively. The ``acceleration'' occurs over 500~AU, again, much
larger than predicted by magneto-hydrodynamical models. This
suggests that the faint and extended H$_2$ emission in B5-IRS~1 is
due to entrainment of surrounding gas by a faster flow. As described
above, Pyo et al. (2003) argue that the entrainment of the LVC in DG
Tau results from interaction with a collimated jet. However, [Fe II]
and H$_2$ observations by Davis et al. (2001, 2003) suggest that
such entrainment produces an H$_2$ velocity similar to, but slightly
lower than that of the [Fe II] emission. Thus, it is more likely
that the faint H$_2$ component of B5-IRS~1 is entrained by a slower
flow component, i.e., a wide-angled component surrounding the
collimated jet. The presence of such an unseen component is indeed
predicted by models (see e.g., Cabrit et al. 1999; Shang et al.
2002) and suggested by observations, including the detection of
H$_2$ emission in cavity walls (Davis et al. 2002; Saucedo et al.
2003).

It is not clear whether the remaining component, i.e., the bright
peak of H$_2$-LVC is also explained by entrainment.
Bacciotti et al. (2002) detect internal flow motion for the LVC in
DG Tau, and their results agree with rotation as predicted by
magneto-centrifugal models. This is corroborated by Takami et al.
(2004), who show that H$_2$ emission in DG Tau is well explained by
an outer extension of the kinematic structure predicted by
magneto-centrifugal models. High-resolution spectro-imaging of the
LVC and proper motion studies could give more crucial constraints on
the nature of this component (e.g., Davis et al. 2006).

\subsection{Episodic Mass Ejection and Optical Outbursts in SVS 13}

HH objects represent fossil records of mass ejection from YSOs. The
spatial structure of blueshifted and redshifted jet knots and bow
shocks often show near-perfect symmetry, indicating that mass
ejection from HH driving sources is episodic (see Reipurth \& Bally
2001, and references therein). Some authors argue that this results
from episodic mass accretion, observed in extreme cases, as FU
Orionis (FUor) or EX Orionis (EXor) outbursts (e.g., Dopita 1978;
Reipurth 1989; Herbig 1989).

To date, a dozen YSOs are known as FUor objects. During an optical
outburst the brightness of the YSO increases by roughly 5 mag on
timescales ranging from less than a year to a decade or more. This
has been successfully modelled as an eruptive accretion event in a
circumstellar disk with an extremely high mass accretion rate
($\sim~10^{-4}$~M$_\odot$\,yr$^{-1}$ --- see e.g., Hartmann \&
Kenyon 1985).
Two scenarios have been proposed to date as a trigger for FUor
outbursts: (1) a runaway instability regulated by the ionization of
hydrogen in the inner disk (Bell \& Lin 1994; Bell et al. 1995), or,
(2) disk perturbations caused by a companion star in a highly
eccentric orbit (Bonnell \& Bastien 1992; Clarke \& Syer 1996).

There is growing evidence that FUor eruptions are a standard, albeit
infrequent phenomenon of the earliest phases of star formation.
Kenyon et al. (1990) show that the number of FUor outbursts observed
within 1 kpc of the Sun in the last 50 yrs is much larger than the
estimated star formation rate within the same region. Muzerolle et
al. (1998) and Calvet et al. (2000) show that the steady disk
accretion rate measured using Br$\gamma$ emission is responsible for
only a tiny fraction (0.01--0.1) of the final stellar mass. In
addition, Muzerolle et al. (1998) suggest that the typical envelope
infall rate onto the accretion disk is 10--100 times larger than the
disk accretion rate. This suggests the presence of episodic and
violent disk accretion onto the star. Based on the results described
above, Kenyon et al. (1990) estimate that each YSO experiences about
10 FUor bursts during its evolution. These most likely occur during
the Class I or early Class II phase as all known FUors are
associated with reflection nebulae analogous to these protostars
(Goodrich 1987), and, submillimetre observations reveal that FUors
have accretion disks comparable in mass to Class I protostars
(Sandell \& Weintraub 2001).

In addition to FUors, there are also a substantial number of stars,
namely ``EX Ori" (EXor) objects, which undergo more modest and
repeatable bursts (Herbig 1989). While the high brightness of FUors
lasts several decades or even more, EXor bursts remain at maximum
for only a few hundred days. In addition, these two kinds of bursts
exhibit remarkably different spectra. Those of FUors do not contain
many emission features and have weak H$\alpha$ emission with very
deep and broad blueshifted absorption (see e.g., Hartmann \& Calvet
1995). In contrast, the spectra of EXors show a rich variety of
emission lines similar to classical T Tauri stars. Even so, it is
likely that both the FUor and EXor bursts result from episodic and
violent disk accretion (see e.g., Hartmann, Kenyon, \& Hartigan
1993).


Among the objects we observed, SVS 13 underwent an optical outburst
about 10 yrs ago, in 1988--1990 (Eisl\"offel et al. 1991; Mauron \&
Touvenot 1991), with $V$ and $R$ magnitudes increasing by up to 3
mag. Optical spectra exhibited bright emission lines even after the
burst, suggesting that the burst is of the EXor variety rather than
FUor (Eisl\"offel et al. 1991), although the burst continued for at
least 500 days (Aspin \& Sandell 1994), longer than the previously
mentioned typical time scale for an EXor burst (Herbig 1989).

The observed outflow structure in SVS 13 could provide an excellent
opportunity to investigate the relationship between optical
outbursts and episodic mass ejection. Figure 2 shows three peaks in
the H$_2$ 2.122~$\mu$m and [Fe II] 1.644~$\mu$m emission of SVS 13.
These are: (1) an emission knot in the H$_2$-HVC, peaking at
1.3\arcsec\, from the object and with a radial velocity of
$\sim~-80$~km\,s$^{-1}$; (2) the H$_2$-LVC, peaking at 0.2\arcsec\,
from the star and with a radial velocity of $\sim~-30$~km\,s$^{-1}$;
(3) the [Fe II] component, which is offset 0.1\arcsec\, from the
star and has a velocity of $-$140~km\,s$^{-1}$. The dynamical ages
of these structures are 30--60, 10--30 and 1--3~yr, respectively,
assuming the same inclination angle to the plane of the sky as the
extended jet (20--40$^\circ$ --- Appendix A). The dynamical age of
the knot in the H$_2$ jet suggests that it predates any episodic mass
accretion/ejection event associated with the 1988-1990 optical burst.
In contrast, the age of the H$_2$ low-velocity component does match
the time of the burst.

Moreover, recent proper motion measurements suggest that the knot in
the H$_2$ jet moves much more slowly than expected from its radial
velocity (Davis et al. 2006). This would suggest an even large
dynamical age than 30--60 yrs.
In any event further measurements are required for
a better determination of the dynamical age, hence enabling a more
detailed investigation of the relation (if any) between the ejection
and optical burst events.

\section{Conclusions}

We detect a number of infrared H$_2$ and [Fe II] lines towards three
Herbig-Haro driving sources: SVS 13,  B5-IRS~1 and HH~34~IRS. These
line include H$_2$ transitions with $v$=1--3 and $J$=1--11, and
several [Fe II] lines with upper energy levels
$E/k=1.1-2.7\times10^4$~K. Outflowing gas in all objects show the
presence of two blueshifted components at high ($-$70 to
$-$130~km\,s$^{-1}$) and low velocities ($-$10 to
$-$30~km\,s$^{-1}$), characteristically similar to forbidden line
emission associated with T Tauri stars but without a low velocity
peak for the forbidden lines, which is observed in the H$_2$
emission. In SVS 13, the HVC also appears in H$_2$ emission. This
reflects the low-excitation nature of the outflowing gas close to
these HH driving sources.

The observed H$_2$ flux ratios indicate a temperature and density of
$2-3\times10^3$~K and $\ge 10^5$~cm$^{-3}$, respectively. In
particular the temperature obtained from the $I_{2-1 S(1)}$/$I_{1-0
S(1)}$ ratio remains quite constant at $2\times10^3$~K, lending
support to shock heating for the excitation of the gas. In the jet
of SVS 13, the gas density is seen to decrease further downstream.
The observed level population of H$_2$ indicates a hydrogen number
density $n_{H}
>10^5 - 10^8$~cm$^{-3}$ depending on whether the excitation is
dominated by H-H$_2$ or H$_2$-H$_2$ collisions, respectively. The
observed electron density and temperature in [Fe II] emission line
regions are $\sim 10^4$~cm$^{-3}$ or greater, similar to FEL regions
associated with T Tauri stars.

The absence of an LVC in [Fe II] emission in all three HH driving
sources matches models based on magneto-centrifugal winds. Our
results in B5-IRS~1 and HH~34~IRS so far support disk-wind models,
which predict that the jet/wind originates from the disk surface at
radii of a few AU. Currently the X-wind model, which predicts that
the jet/wind originates from the inner disk edge within 0.1~AU of
the star, cannot explain the absence of redshifted emission at the
driving source. More detailed information regarding jet launching
could be extracted if accurate collisional rate coefficients for [Fe
II] $^2P$ states are obtained. In B5-IRS~1 the faint extended
component of H$_2$-LVC slowly accelerates over a large spatial scale
($\sim$~20~km\,s$^{-1}$
at $\sim$~500~AU). We suggest that the emission is due to
interaction between a slow and unseen wide-angled wind and the
surrounding ambient gas.

\acknowledgments

We thank an anonymous referee for useful comments. We acknowledge
the data analysis facilities provided by the Starlink Project which
is run by CCLRC on behalf of PPARC. MT thanks PPARC for support
through a PDRA up until September 2004. TPR acknowledges support
from Science Foundation Ireland through contract 04/BR6/PO2741. This
research has made use of the Simbad database, operated at CDS.
Strasbourg, France, and of the NASA's Astrophysics Data System
Abstract Service.



{\it Facilities:} \facility{Subaru (IRCS)}.



\appendix

\section[targets]{Notes on Targets}
\subsection{SVS 13}
SVS 13 is the driving source of the HH 7--11 outflow, which extends
to the southeast for over an arcminute (see e.g. Bachiller et al.
2000). The parent star-forming region, NGC 1333, lies between 220
and 350 pc away (Herbig \& Jones 1983; Cernis 1990; de Zeeuw et al.
1999). For this work, we adopt a distance to SVS 13 of 300 pc.
Analysis of the radial and proper motions of the atomic line
emission from HH 7--10 yield radial velocities of
$40-90$~km\,s$^{-1}$ (Solf \& B\"ohm 1987; Hartigan et al. 1987;
Movsessian et al. 2000) and tangential velocities of 20 -- 60 km
s$^{-1}$ (0.015\arcsec -- 0.044\arcsec\,yr$^{-1}$ --- Noriega-Crespo
\& Garnavich 2001; Khanzadyan et al. 2003). Based on Noriega-Crespo
\& Garnavich (2001) and adopted distance of 300 pc, we estimate an
outflow angle to the line of sight of 20$^\circ$--40$^\circ$. The
inner knot HH 11 possess a much higher radial velocity
($180-200$~km\,s$^{-1}$) than HH 7--10, suggesting that it is
generated by a relatively recent high-speed outburst (Solf \& B\"ohm
1987). SVS 13 exhibited a large increase of its brightness in the
optical and near-IR in 1988--1990 (by 2.5--3.4 and 1--2 mag.,
respectively
--- Eisl\"offel et al. 1991; Mauron \& Thouvenot 1991; Liseau, Lorenzetti
\& Molinari 1992). VLA observations by Anglada (2000) revealed two
components in the E-W direction separated by 0.3\arcsec, indicating
that SVS 13 is a close binary system.

The presence of near-IR H$_2$ emission at SVS 13 itself was first
reported by Carr (1990).
H$_2$ and [Fe II] emission in the extended outflow HH 7--11 have
been observed for decades by a number of authors (Zealey, Williams
\& Sandel 1984, Garden, Russell \& Burton 1990, Stapelfeldt et al.
1991, Carr 1993, Fernandes \& Brand 1995, Gredel 1996; Everett 1997,
Chrysostomou et al. 2000; Davis et al. 2000, 2001, 2002 and
Khanzadyan et al. 2003 for H$_2$; Stapelfeldt et al. 1991 and
Everett 1997 for [Fe II]).

\subsection{B5-IRS~1}
This protostar drives a parsec-scale, jet-like, bipolar HH flow HH
366 which extends in an east-west direction (Bally et al. 1996).
Bally et al. (1996) shows that the outflow extends for about
22\arcmin, corresponding to a projected length of about 2.2 pc, and
H$\alpha$ and [SII] lines reveal radial velocities of $-$30 to
$-$10~km\,s$^{-1}$ and 35 to 90~km\,s$^{-1}$ for the eastern and
western chains respectively.
Yu et al. (1999) derive an inclination angle of 77$^\circ$ to the
line of sight from the projected size and velocity extent of the
associated molecular outflow. The parent dark cloud Barnard 5 is
located at the eastern end of the Perseus molecular cloud complex,
which also include NGC 1333 and SVS 13. We then adopt the same
distance as SVS 13: that is, 300 pc. The presence of H$_2$ and [Fe
II] emission at the driving source was first revealed by Reipurth \&
Aspin (1997) and Davis et al. (2003), respectively.

\subsection{HH~34~IRS}
The HH~34~IRS is responsible for a number of HH objects in an
north-south direction extending over 21\arcmin. These include HH
34S/N/X, HH 40, HH 85--88, HH 126, and HH 173 (Bally \& Devine 1994;
Devine et al. 1997). The object also exhibit a well collimated jet
extending towards the south with a spatial scale of $\sim
20$\arcsec\, (Reipurth et al. 1986, 2000, 2002; B\"uhrke et al.
1988; Ray et al. 1996). Eisl\"offel \& Mundt (1992) and Heathcote \&
Reipurth (1992) derive a inclination angle of 23$^\circ$--28$^\circ$
to the line of the sight, based on their proper motion measurements
and radial velocities in the literature. HH~34~IRS is much less
embedded than the other Class I protostars ($A_V \sim 5$), and its
spectrum at optical wavelengths is rich in emission lines like
classical T Tauri stars (Reipurth et al. 1986).
Estimates of the distance to the parent SFR (Orion OB1 association)
span the range 400--500~pc (see, e.g., Warren \& Hesser 1978; Genzel
\& Stutzki 1989). We adopt a distance of 450 pc in this paper.
Near-IR H$_2$ and [Fe II] emission within a few arcsec of this
object was first observed by Reipurth \& Aspin (1997) and Reipurth
et al. (2000), respectively.

\section[]{Positional Differences between $H$- and $K$-Band Continuum}
Since $H$- and $K$-band spectra were obtained with different
exposures, the spatial structure of emission lines between the two
bands cannot be compared unless their relative position is
determined. Class I protostars are often heavily embedded ($A_V \sim
50$~mag --- Whitney et al. 1997), and we usually see scattered light
even in the $H$- and $K$-bands. This causes a positional difference
between the two bands due to different scattering geometries and
optical depths (see e.g., Weintraub, Kastner \& Whitney 1995). To
determine the relative position of $H$- and $K$-band continuum
emission we obtained exposures in these bands using the imaging mode
of IRCS. The continuum positions for each object were then measured
using other stars in the same field of view. We find that, in
B5-IRS1, the position of the $H$-band continuum is located at $\sim
0.06$\arcsec\, east of the $K$-band continuum, in the same direction
as the blueshifted knot HH 365-E. This trend is what we expect due
to different scattering geometries between these two bands (see
Weintraub et al. 1995). No positional differences exceeding
0.03\arcsec\, (0.5 pixel) are measured in the other objects. Since
the measured positional difference is much smaller than the seeing
(0.6\arcsec), we thus performed comparisons between the two bands
without correcting for these positional differences.




\clearpage




\clearpage

\clearpage

\begin{deluxetable}{ccccrlcr}
\tabletypesize{\scriptsize}
\tablecaption{Target information\label{tbl-targets}}
\tablewidth{0pt}
\tablehead{
\colhead{Object} & \colhead{Distance\tablenotemark{a}} &
\colhead{$V_{\rm LSR}$\tablenotemark{b}} & \colhead{Flow inclination\tablenotemark{c}} &
\colhead{Slit P.A.} & \colhead{Orientation} & \colhead{Band}&
\colhead{Total on-source integration\tablenotemark{d}} \\
& \colhead{(pc)}     & \colhead{(km s$^{-1}$)} & \colhead{(deg.)} & \colhead{(deg.)}
&     &      & (s)
}
\startdata
SVS 13            & 300      &    +8~~      &20--40& 159 &along the jet& $H$ & 1440\\
                  &          &               & & 159      &along the jet& $K$ & ~480\\
                  &          &               & & ~69      &perp. to the jet& $K$ & ~480\\
B5 IRS1           & 300      &    +10~~      &$\sim$80&~73&along the jet& $H$ & 1440\\
                  &          &               & & ~73      &along the jet& $K$ & 1200\\
HH~34~IRS         & 450      &    +8.5      &23--28& 167      &along the jet& $H$ & 1440\\
                  &          &               & & 167      &along the jet& $K$ & 1200\\
\enddata
\tablenotetext{a}{See text.} \tablenotetext{b}{The Local Standard of
Rest velocity of the parent cloud measured by Knee \& Sandel (2000),
Yu et al. (1999), and Chernin \& Masson (1995).}
\tablenotetext{c}{To the line of sight. See Appendix A for details.}
\tablenotetext{d}{Including those with the opposite slit angle
(i.e., P.A. + 180$^\circ$).}
\end{deluxetable}

\begin{table*}
 \caption{H$_2$ fluxes normalized to the 1--0 S(1)
 flux\label{tbl-H2ratios}}
 \begin{center}
{\scriptsize \hspace*{-1cm}
\begin{tabular}{@{}lccccccccc@{}}
Object & Region & 1--0 S(9) & 1--0 S(8) & 1--0 S(7) & 2--1 S(4) & 1--0 S(2) & 2--1 S(3) & 1--0 S(1) & 2--1 S(2) \\
&&1.688~$\mu$m &1.715~$\mu$m &1.748~$\mu$m &2.004~$\mu$m &2.034
$\mu$m &2.073~$\mu$m &2.122~$\mu$m &2.154~$\mu$m \\ \hline
SVS 13 &  1  & 0.02$\pm$0.01& 0.01$\pm$0.01& 0.13$\pm$0.02& $<$0.06    & 0.29$\pm$0.01& 0.07$\pm$0.01& 1.00$\pm$0.01& 0.03$\pm$0.01\\
       &  2  & $<$0.04    & $<$0.02    & 0.11$\pm$0.09& $<$0.07    & 0.23$\pm$0.13& $<$0.13    & 1.00$\pm$0.02& 0.05$\pm$0.02\\
       &  3  & 0.05$\pm$0.01& 0.03$\pm$0.01& 0.12$\pm$0.04& $<$0.04    & 0.33$\pm$0.05& 0.08$\pm$0.06& 1.00$\pm$0.01& $<$0.02    \\
B5-IRS~1 &1  & $<$0.01    & $<$0.01    & 0.15$\pm$0.02& $<$0.04    & 0.34$\pm$0.02& 0.10$\pm$0.01& 1.00$\pm$0.01& 0.04$\pm$0.01\\
         &2  & $<$0.06    & $<$0.09    & 0.34$\pm$0.10& $<$0.43    & 0.38$\pm$0.10& 0.16$\pm$0.07& 1.00$\pm$0.08& 0.15$\pm$0.07\\
HH~34~IRS&1  & $<$0.06    & $<$0.06    & 0.08$\pm$0.06& $<$0.12 &
0.35$\pm$0.03& 0.05$\pm$0.03& 1.00$\pm$0.03& 0.09$\pm$0.03
\\[10pt]
\end{tabular}
\begin{tabular}{@{}ccccccccc@{}}
3--2 S(3) & 1--0 S(0) & 2--1 S(1) & 3--2 S(2) & 3--2 S(1) & 1--0 Q(1) & 1--0 Q(2) & 1--0 Q(3) & 1--0 Q(4)  \\
2.201~$\mu$m  & 2.223~$\mu$m &2.248~$\mu$m &2.287~$\mu$m &2.386
$\mu$m &2.407~$\mu$m &2.413~$\mu$m &2.424~$\mu$m &2.437~$\mu$m \\
\hline
$<$0.02    & 0.24$\pm$0.01& 0.08$\pm$0.01& $<$0.01& $<$0.03 & 0.89$\pm$0.02& 0.25$\pm$0.02& 0.79$\pm$0.02& 0.25$\pm$0.01\\
$<$0.02    & 0.21$\pm$0.02& 0.11$\pm$0.07&  ---   &  ---    & 0.85$\pm$0.47&  ---       & 0.92$\pm$0.35& 0.71$\pm$0.48\\
0.03$\pm$0.01& 0.22$\pm$0.01& 0.12$\pm$0.03&  ---   &  ---    & 0.61$\pm$0.22&  ---       & 1.22$\pm$0.15& 0.42$\pm$0.04\\
$<$0.02    & 0.32$\pm$0.01& 0.11$\pm$0.01& $<$0.02& $<$0.03 & 1.12$\pm$0.03& 0.35$\pm$0.02& 1.06$\pm$0.02& 0.31$\pm$0.02\\
$<$0.07    & 0.24$\pm$0.05& 0.09$\pm$0.05& $<$0.07& $<$0.12 & 1.19$\pm$0.14& 0.08$\pm$0.08& 0.92$\pm$0.12& 0.30$\pm$0.09\\
$<$0.05    & 0.27$\pm$0.02& 0.08$\pm$0.02& $<$0.03& $<$0.10 &
1.04$\pm$0.06& 0.32$\pm$0.10& 0.88$\pm$0.05& 0.27$\pm$0.04
\end{tabular}
}
\end{center}
\end{table*}

\begin{table*}
 \caption{[Fe II] fluxes normalized to the 1.644~$\mu$m
 flux.\label{tbl-FeIIratios}}
\hspace*{-1cm}
\begin{center}
{\scriptsize
\begin{tabular}{@{}lccccccccc@{}}
Object & Region &  $A_V$
                &  $^4D_{5/2}$$\rightarrow$$^4F_{9/2}$
                &  $^4D_{3/2}$$\rightarrow$$^4F_{7/2}$
                &  $^4D_{7/2}$$\rightarrow$$^4F_{9/2}$
                &  $^4D_{1/2}$$\rightarrow$$^4F_{5/2}$
                &  $^4D_{3/2}$$\rightarrow$$^4F_{5/2}$
                &  $^4D_{1/2}$$\rightarrow$$^4F_{3/2}$ \\
& & & 1.533~$\mu$m & 1.600~$\mu$m & 1.644~$\mu$m & 1.664~$\mu$m &
1.712~$\mu$m & 1.745~$\mu$m \\

& & & ($E_u$=1.04 eV)\tablenotemark{a}
& (1.08 eV) & (0.99 eV) & (1.10 eV) & (1.08 eV) & (1.10 eV) \\
\hline

SVS 13 &  1  &  0 & 0.37$\pm$0.02 &  0.31$\pm$0.02 &  1.00$\pm$0.02 &  0.13$\pm$0.01 &  ---         &  ---         \\
       &     & 20 & 0.54$\pm$0.03 &  0.36$\pm$0.02 &  1.00$\pm$0.02 &  0.12$\pm$0.01 &  ---         &  ---         \\ [2pt]
B5-IRS~1 &1  &  0 & 0.23$\pm$0.03 &  0.20$\pm$0.02 &  1.00$\pm$0.02 &  0.13$\pm$0.02 &  0.08$\pm$0.01 &  $<$0.05     \\
         &   & 20 & 0.34$\pm$0.04 &  0.23$\pm$0.02 &  1.00$\pm$0.02 &  0.12$\pm$0.02 &  0.06$\pm$0.01 &  $<$0.03     \\ [2pt]
         &2  &  0 & 0.17$\pm$0.02 &  0.21$\pm$0.02 &  1.00$\pm$0.02 &  0.15$\pm$0.02 &  0.08$\pm$0.02 &  0.04$\pm$0.03 \\
         &   & 20 & 0.25$\pm$0.03 &  0.25$\pm$0.02 &  1.00$\pm$0.02 &  0.14$\pm$0.02 &  0.07$\pm$0.01 &  0.03$\pm$0.02 \\ [2pt]
         &3  &  0 & 0.22$\pm$0.03 &  0.23$\pm$0.03 &  1.00$\pm$0.03 &  0.24$\pm$0.03 &  0.10$\pm$0.03 &  0.07$\pm$0.04 \\
         &   & 20 & 0.32$\pm$0.05 &  0.26$\pm$0.03 &  1.00$\pm$0.03 &  0.23$\pm$0.03 &  0.08$\pm$0.02 &  0.05$\pm$0.03 \\ [2pt]
     &4+5&  0 & ---         &  0.13$\pm$0.04 &  1.00$\pm$0.05 &  0.09$\pm$0.05 &  0.06$\pm$0.05 &  $<$0.06     \\
         &   & 20 & ---         &  0.15$\pm$0.05 &  1.00$\pm$0.05 &  0.09$\pm$0.05 &  0.05$\pm$0.04 &  $<$0.05     \\ [2pt]
HH~34~IRS &1  &  0 & 0.32$\pm$0.01 &  0.20$\pm$0.01 &  1.00$\pm$0.01 &  0.12$\pm$0.01 &  0.09$\pm$0.01 &  0.06$\pm$0.01\\
          &   & 30 & 0.56$\pm$0.02 &  0.25$\pm$0.01 &  1.00$\pm$0.01 &  0.11$\pm$0.01 &  0.07$\pm$0.01 &  0.04$\pm$0.01\\ [2pt]
          &2  &  0 & 0.28$\pm$0.01 &  0.21$\pm$0.01 &  1.00$\pm$0.01 &  0.13$\pm$0.01 &  0.07$\pm$0.01 &  0.07$\pm$0.01\\
          &   & 15 & 0.37$\pm$0.01 &  0.24$\pm$0.01 &  1.00$\pm$0.01 &  0.13$\pm$0.01 &  0.06$\pm$0.01 &  0.06$\pm$0.01\\ [2pt]
          &3  &  0 & 0.21$\pm$0.01 &  0.15$\pm$0.01 &  1.00$\pm$0.01 &  0.10$\pm$0.01 &  0.05$\pm$0.01 &  0.04$\pm$0.01\\
          &   & 15 & 0.28$\pm$0.01 &  0.17$\pm$0.01 &  1.00$\pm$0.01 &  0.09$\pm$0.01 &  0.04$\pm$0.01 &  0.03$\pm$0.01\\ [2pt]
          &4  &  0 & ---     &  0.11$\pm$0.01 &  1.00$\pm$0.01 &  0.09$\pm$0.01 &  0.04$\pm$0.01 &  0.05$\pm$0.01 \\
          &   & 15 & ---     &  0.12$\pm$0.01 &  1.00$\pm$0.01 &  0.08$\pm$0.01 &  0.03$\pm$0.01 &  0.04$\pm$0.01 \\ [3pt]
\end{tabular}
\begin{tabular}{@{}lccccccccc@{}}
Object & Region &  $A_V$
                &  $^4P_{3/2}$$\rightarrow$$^4D_{7/2}$
                &  $^2P_{1/2}$$\rightarrow$$^4P_{1/2}$
                &  $^2P_{3/2}$$\rightarrow$$^4P_{5/2}$
                &  $^4P_{3/2}$$\rightarrow$$^4D_{1/2}$
                &  $^2P_{3/2}$$\rightarrow$$^4P_{3/2}$
                &  $^2P_{3/2}$$\rightarrow$$^4P_{1/2}$ \\
& & & 1.749~$\mu$m & 2.007~$\mu$m & 2.047~$\mu$m & 2.072~$\mu$m &
2.133~$\mu$m & 2.244~$\mu$m \\

& & & ($E_u$=1.70 eV)$^a$ & (2.34 eV) & (2.28 eV) & (1.70 eV) &
(2.28 eV) & (2.28 eV) \\ \hline

SVS 13 &  1  &  0 &  $<$0.14     &  0.40$\pm$0.31 &  $<$0.23     & $<$0.14      &  0.11$\pm$0.05 &  $<$0.11       \\
       &     & 20 &  $<$0.10     &  0.17$\pm$0.13 &  $<$0.09     & $<$0.05      &  0.04$\pm$0.02 &  $<$0.03       \\[2pt]
B5-IRS~1 &1  &  0 &  $<$0.02     &  $<$0.08     & 0.026$\pm$0.018& $<$0.03      & 0.030$\pm$0.016&  $<$0.02       \\
         &   & 20 &  $<$0.02     &  $<$0.03     & 0.010$\pm$0.007& $<$0.01      & 0.010$\pm$0.005&  $<$0.007      \\[2pt]
         &2  &  0 &  ---         &  $<$0.08     & 0.025$\pm$0.018& 0.041$\pm$0.017& 0.047$\pm$0.017&  $<$0.01       \\
         &   & 20 &  ---         &  $<$0.03     & 0.010$\pm$0.007& 0.015$\pm$0.006& 0.016$\pm$0.006&  $<$0.004      \\[2pt]
         &3  &  0 &  ---         &  $<$0.17     &  $<$0.02     &  0.03$\pm$0.03 &  $<$0.02     &  $<$0.06       \\
         &   & 20 &  ---         &  $<$0.07     &  $<$0.009    &  0.01$\pm$0.01 &  $<$0.007    &  $<$0.02       \\[2pt]
     &4+5&  0 &  $<$0.11     &  $<$0.11     &  $<$0.06     &  $<$0.04     &  $<$0.05     &  $<$0.04       \\
         &   & 20 &  $<$0.08     &  $<$0.05     &  $<$0.02     &  $<$0.02     &  $<$0.02     &  $<$0.01       \\[2pt]
HH~34~IRS &1  &  0 &  $<$0.015   &  $<$0.03     & 0.066$\pm$0.005&  $<$0.007    & 0.028$\pm$0.004&  0.017$\pm$0.003 \\
          &   & 30 &  $<$0.009   &  $<$0.007    & 0.016$\pm$0.001&  $<$0.002    & 0.006$\pm$0.001&  0.003$\pm$0.001 \\[2pt]
          &2  &  0 &  $<$0.005   &  $<$0.08     & 0.058$\pm$0.003&  $<$0.005    &0.032$\pm$0.003 &  0.015$\pm$0.002 \\
          &   & 15 &  $<$0.004   &  $<$0.04     & 0.028$\pm$0.002&  $<$0.002    &0.014$\pm$0.001 &  0.006$\pm$0.001 \\[2pt]
          &3  &  0 &  $<$0.012   &  $<$0.03     & $<$0.004     &  $<$0.005    & 0.008$\pm$0.004&  $<$0.002      \\
          &   & 15 &  $<$0.009   &  $<$0.01     & $<$0.002     &  $<$0.002    & 0.004$\pm$0.002&  $<$0.001      \\[2pt]
          &4  &  0 & 0.011$\pm$0.010&  ---         & $<$0.005     &  $<$0.005    & 0.010$\pm$0.004&  $<$0.005      \\
          &   & 15 & 0.009$\pm$0.008&  ---         & $<$0.003     &  $<$0.002    & 0.004$\pm$0.002&  $<$0.002
\end{tabular}
}
\end{center}
\tablenotetext{a}{Upper energy level from the ground state.}
\end{table*}

\begin{deluxetable}{lccccc}
\tabletypesize{\scriptsize} \tablecaption{Estimated extinction and
excitation temperatures of H$_2$ emission line
regions.\label{tbl-H2phys}} \tablewidth{0pt} \tablehead{
\colhead{Object} & \colhead{Region} & \multicolumn{3}{c}{$A_V$} & \colhead{$T_{2-1S(1)/1-0S(1)}$}\\
& & \colhead{S(1)/Q(3)} & \colhead{S(0)/Q(2)} & \colhead{S(2)/Q(4)}&
\colhead{(10$^3$ K)} }
\startdata
SVS 13 &  1  & 6.1$\pm$0.8 & $<$1      & 15$\pm$2  &1.92$\pm$0.06\\
       &  2  & ---       & ---       & ---     &2.1~~$\pm$0.5~~\\
       &  3  &  28$\pm$6   & ---       & 30$\pm$6  &2.1~~$\pm$0.2~~\\
B5-IRS~1 &1  &  19$\pm$1   & $<$6      & 16$\pm$3  &2.05$\pm$0.06\\
         &2  &  14$\pm$6   & ---       & 16$\pm$14 &1.9~~$\pm$0.4~~\\
HH~34~IRS&1  &  12$\pm$3   & $<$40     & 12$\pm$6 &1.9~~$\pm$0.1~~
\enddata
\end{deluxetable}

\begin{deluxetable}{lccc}
\tabletypesize{\scriptsize} \tablecaption{Estimated extinction in
[Fe II] emission line regions.\label{tbl-FeIIphys}}
 \tablewidth{0pt}
\tablehead{
\colhead{Object} & \colhead{Region} & \multicolumn{2}{c}{$A_V$} \\
& & \colhead{$I_{1.712}/I_{1.600}$} & \colhead{$I_{1.745}/I_{1.664}$}
}
\startdata
SVS 13 &  1   & ---      & --- \\
B5-IRS~1 &1   & 20$\pm$10& --- \\
         &2   & 20$\pm$10& --- \\
         &3   & 30$\pm$20& --- \\
         &4+5 &  $<$50   & --- \\
HH~34~IRS &1  & 32$\pm$6 & $<$8  \\
          &2  & 15$\pm$5 & $<$14 \\
          &3  & 15$\pm$7 & $<$8  \\
          &4  & $<$30    & $<$30
\enddata
\end{deluxetable}






\end{document}